\documentclass[sigconf]{acmart}

\usepackage{bm}
\usepackage{amsmath}
\usepackage{amssymb}
\usepackage{caption}
\usepackage{subcaption}
\usepackage{graphicx}
\usepackage{algorithm}
\usepackage{algcompatible}
\usepackage{multirow}

\hypersetup{colorlinks=true,urlcolor=blue}

\newtheorem{theorem}{Theorem}
\newtheorem{proposition}{Proposition}

\copyrightyear{2019}
\acmYear{2019} 
\setcopyright{none}
\acmConference[WWW 2019]{The 2019 Web Conference}{May 13--17, 2019}{San Francisco, USA}

\begin{document}

\title{Multi-Task Feature Learning for Knowledge Graph Enhanced Recommendation}

\author{Hongwei Wang$^{1,2}$, Fuzheng Zhang$^3$, Miao Zhao$^4$, Wenjie Li$^4$, Xing Xie$^2$, Minyi Guo$^1$}
\authornote{M. Guo is the corresponding author. This work was partially sponsored by the National Basic Research 973 Program of China under Grant 2015CB352403.}
\affiliation{$^1$Shanghai Jiao Tong University, Shanghai, China}
\affiliation{$^2$Microsoft Research Asia, Beijing, China}
\affiliation{$^3$Meituan-Dianping Group, Beijing, China}
\affiliation{$^4$The Hong Kong Polytechnic University, Hong Kong, China}
\email{wanghongwei55@gmail.com, zhangfuzheng@meituan.com, {csmiaozhao, cswjli}@comp.polyu.edu.hk}
\email{xingx@microsoft.com, guo-my@cs.sjtu.edu.cn}

\begin{abstract}
	Collaborative filtering often suffers from sparsity and cold start problems in real recommendation scenarios, therefore, researchers and engineers usually use side information to address the issues and improve the performance of recommender systems.
	In this paper, we consider knowledge graphs as the source of side information.
 	We propose \textbf{MKR}, a \textbf{M}ulti-task feature learning approach for \textbf{K}nowledge graph enhanced \textbf{R}ecommendation.
 	MKR is a deep end-to-end framework that utilizes knowledge graph embedding task to assist recommendation task.
 	The two tasks are associated by cross$\&$compress units, which automatically share latent features and learn high-order interactions between items in recommender systems and entities in the knowledge graph.
 	We prove that cross$\&$compress units have sufficient capability of polynomial approximation, and show that MKR is a generalized framework over several representative methods of recommender systems and multi-task learning.
 	Through extensive experiments on real-world datasets, we demonstrate that MKR achieves substantial gains in movie, book, music, and news recommendation, over state-of-the-art baselines.
 	MKR is also shown to be able to maintain a decent performance even if user-item interactions are sparse.
\end{abstract}


\maketitle

\subsection*{\small{ACM Reference Format:}}
\vspace{-0.05in}
	{\small
		Hongwei Wang, Fuzheng Zhang, Miao Zhao, Wenjie Li, Xing Xie, and Minyi Guo.
		2019.
		Multi-Task Feature Learning for Knowledge Graph Enhanced Recommendation
		In \textit{Proceedings of The 2019 Web Conference (WWW 2019)}.
		ACM, New York, NY, USA, 11 pages.
		https://doi.org/xxxxx
	}

\section{Introduction}
	Recommender systems (RS) aims to address the information explosion and meet users personalized interests.
	One of the most popular recommendation techniques is collaborative filtering (CF) \cite{koren2009matrix}, which utilizes users' historical interactions and makes recommendations based on their common preferences.
	However, CF-based methods usually suffer from the sparsity of user-item interactions and the cold start problem.
	Therefore, researchers propose using \textit{side information} in recommender systems, including social networks \cite{jamali2010matrix}, attributes \cite{wang2018shine}, and multimedia (e.g., texts \cite{wang2015collaborative}, images \cite{zhang2016collaborative}).
	\textit{Knowledge graphs} (KGs) are one type of side information for RS, which usually contain fruitful facts and connections about items.
	Recently, researchers have proposed several academic and commercial KGs, such as NELL\footnote{\url{http://rtw.ml.cmu.edu/rtw/}}, DBpedia\footnote{\url{http://wiki.dbpedia.org/}}, Google Knowledge Graph\footnote{\url{https://developers.google.com/knowledge-graph/}} and Microsoft Satori\footnote{\url{https://searchengineland.com/library/bing/bing-satori}}.
	Due to its high dimensionality and heterogeneity, a KG is usually pre-processed by \textit{knowledge graph embedding} (KGE) methods \cite{wang2018graphgan}, which embeds entities and relations into low-dimensional vector spaces while preserving its inherent structure.
	
	\vspace{0.5em}
	\noindent\textbf{Existing KG-aware methods}
	
	Inspired by the success of applying KG in a wide variety of tasks, researchers have recently tried to utilize KG to improve the performance of recommender systems \cite{yu2014personalized,zhao2017meta,wang2018dkn,wang2018ripple,zhang2016collaborative}.
	Personalized Entity Recommendation (PER) \cite{yu2014personalized} and Factorization Machine with Group lasso (FMG) \cite{zhao2017meta} treat KG as a heterogeneous information network, and extract meta-path/meta-graph based latent features to represent the connectivity between users and items along different types of relation paths/graphs.
	It should be noted that PER and FMG rely heavily on manually designed meta-paths/meta-graphs, which limits its application in generic recommendation scenarios.
	Deep Knowledge-aware Network (DKN) \cite{wang2018dkn} designs a CNN framework to combine entity embeddings with word embeddings for news recommendation.
	However, the entity embeddings are required in advance of using DKN, causing DKN to lack an end-to-end way of training.
	Another concern about DKN is that it can hardly incorporate side information other than texts.
	RippleNet \cite{wang2018ripple} is a memory-network-like model that propagates users' potential preferences in the KG and explores their hierarchical interests.
	But the importance of relations is weakly characterized in RippleNet, because the embedding matrix of a relation $\bf R$ can hardly be trained to capture the sense of importance in the quadratic form ${\bf v}^\top {\bf R} {\bf h}$ ($\bf v$ and $\bf h$ are embedding vectors of two entities).
	%Moreover, RippleNet is sensitive to the density of user-item interactions (shown in the experiments section), which limits its usability in sparse scenarios.
	%Moreover, the size of ripple set may go unpredictably with the increase of the size of KG, which incurs heavy computation and storage overhead.
	Collaborative Knowledge base Embedding (CKE) \cite{zhang2016collaborative} combines CF with structural knowledge, textual knowledge, and visual knowledge in a unified framework.
	However, the KGE module in CKE (i.e., TransR \cite{lin2015learning}) is more suitable for in-graph applications (such as KG completion and link prediction) rather than recommendation.
	In addition, the CF module and the KGE module are loosely coupled in CKE under a Bayesian framework, making the supervision from KG less obvious for recommender systems.
	
	\vspace{0.5em}
	\noindent\textbf{The proposed approach}	
	
	To address the limitations of previous work, we propose MKR, a multi-task learning (MTL) approach for knowledge graph enhanced recommendation.
	MKR is a generic, end-to-end deep recommendation framework, which aims to utilize KGE task to assist recommendation task\footnote{KGE task can also benefit from recommendation task empirically as shown in the experiments section.}.
	Note that the two tasks are not mutually independent, but are highly correlated since an item in RS may associate with one or more entities in KG.
	Therefore, an item and its corresponding entity are likely to have a similar proximity structure in RS and KG, and share similar features in low-level and non-task-specific latent feature spaces \cite{long2017learning}.
	We will further validate the similarity in the experiments section.
	To model the shared features between items and entities, we design a \textit{cross$\&$compress unit} in MKR.
	The cross$\&$compress unit explicitly models high-order interactions between item and entity features, and automatically control the cross knowledge transfer for both tasks.
	Through cross$\&$compress units, representations of items and entities can complement each other, assisting both tasks in avoiding fitting noises and improving generalization.
	The whole framework can be trained by alternately optimizing the two tasks with different
frequencies, which endows MKR with high flexibility and adaptability in real recommendation scenarios.
	
	We probe the expressive capability of MKR and show, through theoretical analysis, that the cross$\&$compress unit is capable of approximating sufficiently high order feature interactions between items and entities.
	We also show that MKR is a generalized framework over several representative methods of recommender systems and multi-task learning, including factorization machines \cite{rendle2010factorization, rendle2012factorization}, deep$\&$cross network \cite{wang2017deep}, and cross-stitch network \cite{misra2016cross}.
	Empirically, we evaluate our method in four recommendation scenarios, i.e., movie, book, music, and news recommendations.
	The results demonstrate that MKR achieves substantial gains over state-of-the-art baselines in both click-through rate (CTR) prediction (e.g., $11.6\%$ $AUC$ improvements on average for movies) and top-$K$ recommendation (e.g., $66.4\%$ $Recall@10$ improvements on average for books).
	MKR can also maintain a decent performance in sparse scenarios.
	
	\vspace{0.5em}
	\noindent\textbf{Contribution}		
	
	It is worth noticing that the problem studied in this paper can also be modelled as \textit{cross-domain recommendation} \cite{tang2012cross} or \textit{transfer learning} \cite{pan2010survey}, since we care more about the performance of recommendation task.
	However, the key observation is that though cross-domain recommendation and transfer learning have single objective for the target domain, their loss functions still contain constraint terms for measuring data distribution in the source domain or similarity between two domains.
	In our proposed MKR, the KGE task serves as the constraint term \textit{explicitly} to provide regularization for recommender systems.
	We would like to emphasize that the major contribution of this paper is exactly modeling the problem as multi-task learning: We go a step further than cross-domain recommendation and transfer learning by finding that the inter-task similarity is helpful to not only recommender systems but also knowledge graph embedding, as shown in theoretical analysis and experiment results.

\section{Our Approach}
	In this section, we first formulate the knowledge graph enhanced recommendation problem, then introduce the framework of MKR and present the design of the cross$\&$compress unit, recommendation module and KGE module in detail.
	We lastly discuss the learning algorithm for MKR.
	
	\begin{figure*}[t]
		\centering
        \begin{subfigure}[b]{0.5\textwidth}
            \includegraphics[width=\textwidth]{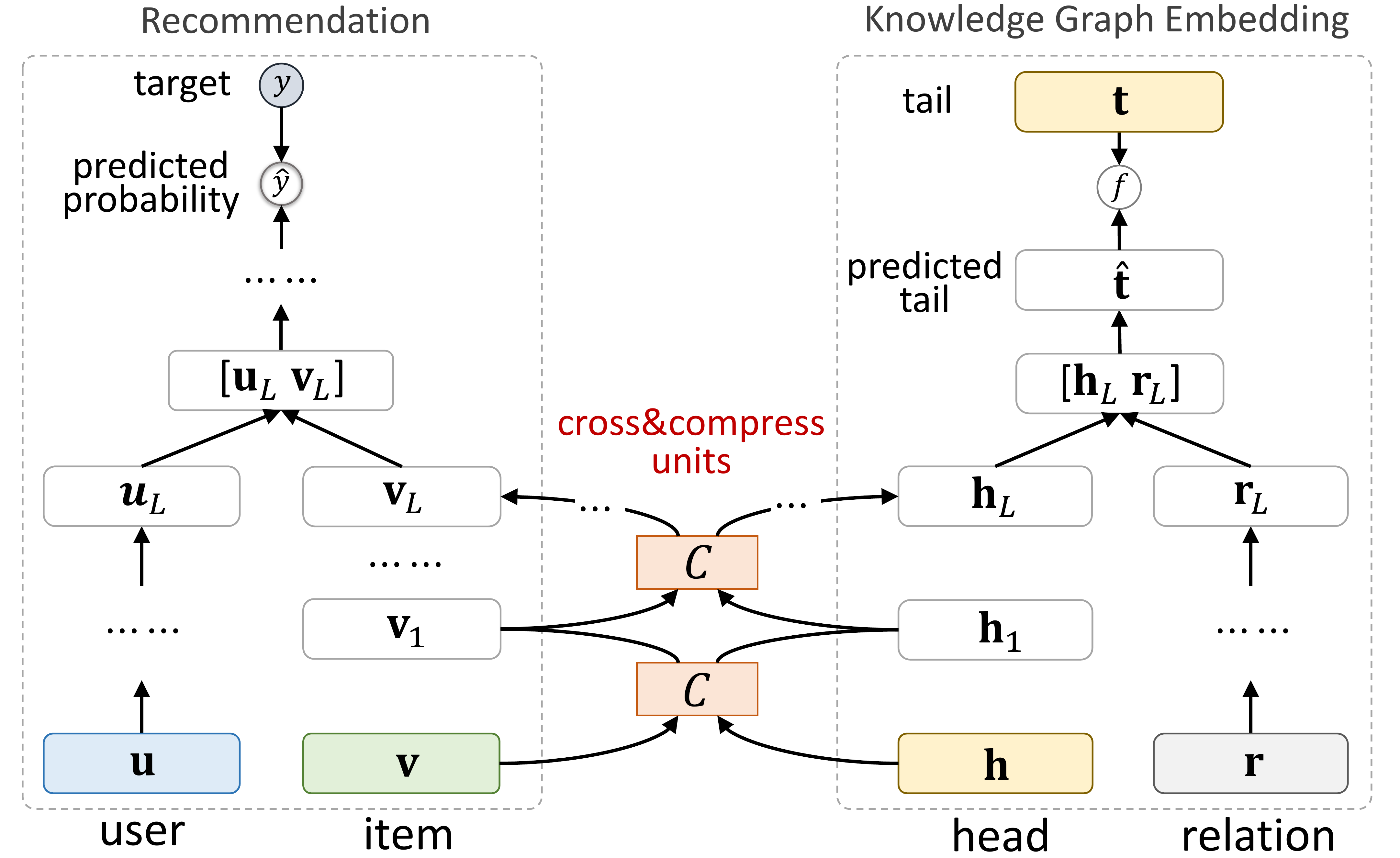}
            \caption{Framework of MKR}
            \label{fig:framework}
        \end{subfigure}
        \hspace{0.06\textwidth}
        \begin{subfigure}[b]{0.36\textwidth}
            \includegraphics[width=\textwidth]{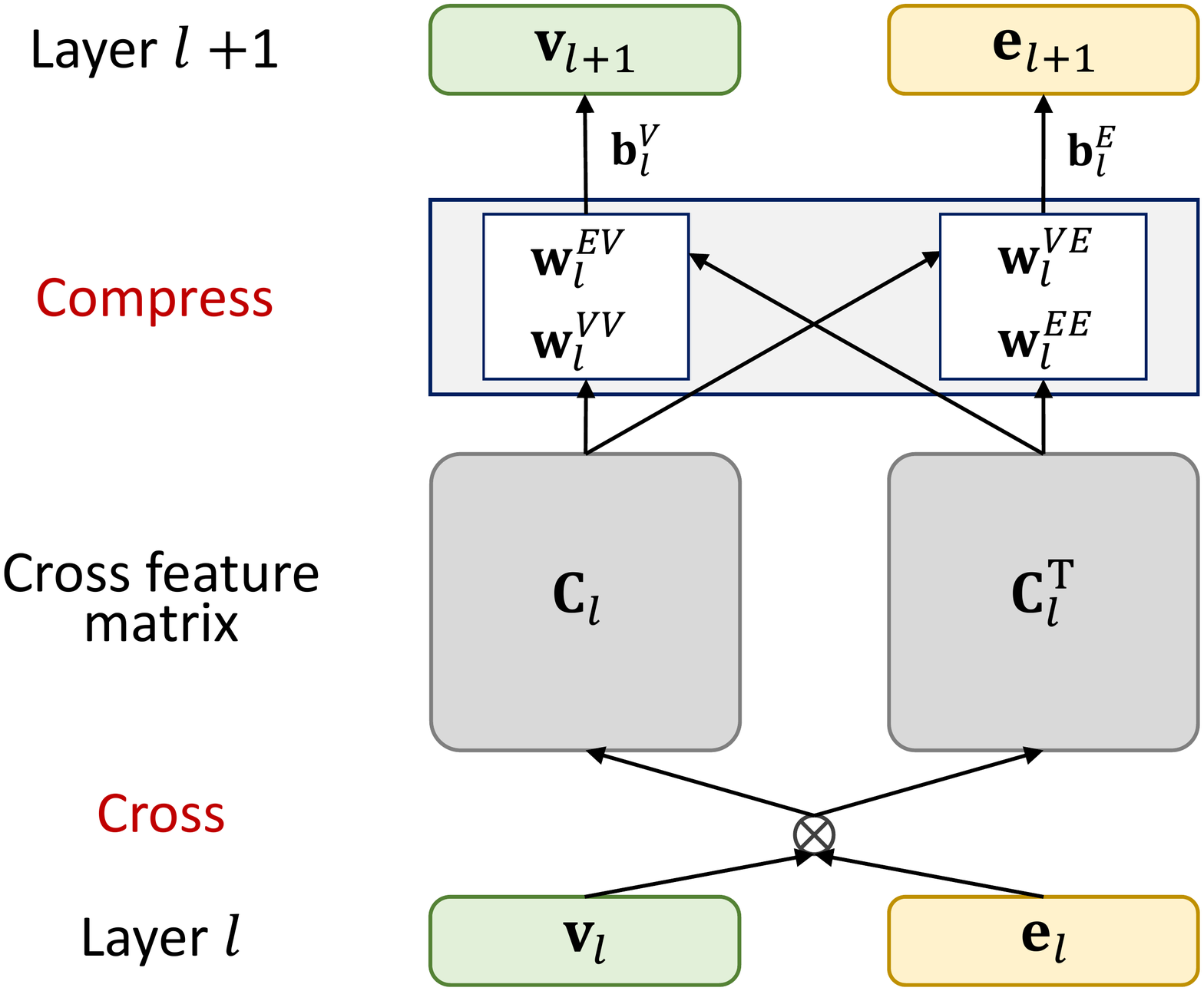}
            \caption{Cross$\&$compress unit}
            \label{fig:cross_feature_sharing_unit}
        \end{subfigure}
        \caption{(a) The framework of MKR. The left and right part illustrate the recommendation module and the KGE module, respectively, which are bridged by the cross$\&$compress units. (b) Illustration of a cross$\&$compress unit. The cross$\&$compress unit generates a cross feature matrix from item and entity vectors by \textit{cross} operation, and outputs their vectors for the next layer by \textit{compress} operation.}
        \label{fig:1}
    \end{figure*}
    
   \subsection{Problem Formulation}
		We formulate the knowledge graph enhanced recommendation problem in this paper as follows.
		In a typical recommendation scenario, we have a set of $M$ users $\mathcal U = \{u_1, u_2, ..., u_M\}$ and a set of $N$ items $\mathcal V = \{v_1, v_2, ..., v_N\}$.
		The user-item interaction matrix ${\bf Y} \in \mathbb R^{M \times N}$ is defined according to users' implicit feedback, where $y_{uv} = 1$ indicates that user $u$ engaged with item $v$, such as behaviors of clicking, watching, browsing, or purchasing; otherwise $y_{uv} = 0$.
		Additionally, we also have access to a knowledge graph $\mathcal G$, which is comprised of entity-relation-entity triples $(h, r, t)$.
		Here $h$, $r$, and $t$ denote the head, relation, and tail of a knowledge triple, respectively.
		For example, the triple (\textit{Quentin Tarantino}, \textit{film.director.film}, \textit{Pulp Fiction}) states the fact that Quentin Tarantino directs the film Pulp Fiction.
		In many recommendation scenarios, an item $v \in \mathcal V$ may associate with one or more entities in $\mathcal G$.
		For example, in movie recommendation, the item "Pulp Fiction" is linked with its namesake in a KG, while in news recommendation, news with the title "Trump pledges aid to Silicon Valley during tech meeting" is linked with entities "Donald Trump" and "Silicon Valley" in a KG.
		
		Given the user-item interaction matrix $\bf Y$ as well as the knowledge graph $\mathcal G$, we aim to predict whether user $u$ has potential interest in item $v$ with which he has had no interaction before.
		Our goal is to learn a prediction function ${\hat y}_{uv} = \mathcal F(u, v | \Theta, \bf Y, \mathcal G)$, where ${\hat y}_{uv}$ denotes the probability that user $u$ will engage with item $v$, and $\Theta$ is the model parameters of function $\mathcal F$.

	\subsection{Framework}		
		The framework of MKR is illustrated in Figure \ref{fig:framework}.
		MKR consists of three main components: recommendation module, KGE module, and cross$\&$compress units.
		(1) The recommendation module on the left takes a user and an item as input, and uses a multi-layer perceptron (MLP) and cross$\&$compress units to extract short and dense features for the user and the item, respectively.
		The extracted features are then fed into another MLP together to output the predicted probability.
		(2) Similar to the left part, the KGE module in the right part also uses multiple layers to extract features from the head and relation of a knowledge triple, and outputs the representation of the predicted tail under the supervision of a score function $f$ and the real tail.
		(3) The recommendation module and the KGE module are bridged by specially designed cross$\&$compress units.
		The proposed unit can automatically learn high-order feature interactions of items in recommender systems and entities in the KG.

	\subsection{Cross$\&$compress Unit}
		%In MKR, recommendation and KGE are highly correlated, because an item in recommender systems may associate with one or more entities in the KG.
		To model feature interactions between items and entities, we design a cross$\&$compress unit in MKR framework.
		As shown in Figure \ref{fig:cross_feature_sharing_unit}, for item $v$ and one of its associated entities $e$, we first construct $d \times d$ pairwise interactions of their latent feature ${\bf v}_l \in \mathbb R^d$ and ${\bf e}_l \in \mathbb R^d$ from layer $l$:
		\begin{equation}
		\label{eq:cross}
			{\bf C}_l = {\bf v}_l {\bf e}_l^\top =
			\begin{bmatrix}
   				v_l^{(1)} e_l^{(1)} & \cdots & v_l^{(1)} e_l^{(d)}\\
   				\cdots & & \cdots\\
   				v_l^{(d)} e_l^{(1)} & \cdots & v_l^{(d)} e_l^{(d)}
  			\end{bmatrix},
		\end{equation}
		where ${\bf C}_l \in \mathbb R^{d \times d}$ is the cross feature matrix of layer $l$, and $d$ is the dimension of hidden layers.
		This is called the \textit{cross} operation, since each possible feature interaction $v_l^{(i)} e_l^{(j)}, \forall (i, j) \in \{1, ..., d\}^2$ between item $v$ and its associated entity $e$ is modeled explicitly in the cross feature matrix.
		We then output the feature vectors of items and entities for the next layer by projecting the cross feature matrix into their latent representation spaces:
		\begin{equation}
		\label{eq:compress}
		\begin{split}
			{\bf v}_{l+1} =& {\bf C}_l {\bf w}_l^{VV} + {\bf C}_l^\top {\bf w}_l^{EV} + {\bf b}_l^V = {\bf v}_l {\bf e}_l^\top {\bf w}_l^{VV} + {\bf e}_l {\bf v}_l^\top {\bf w}_l^{EV} + {\bf b}_l^V,\\
			{\bf e}_{l+1} =& {\bf C}_l {\bf w}_l^{VE} + {\bf C}_l^\top {\bf w}_l^{EE} + {\bf b}_l^E = {\bf v}_l {\bf e}_l^\top {\bf w}_l^{VE} + {\bf e}_l {\bf v}_l^\top {\bf w}_l^{EE} + {\bf b}_l^E,
		\end{split}
		\end{equation}
		where ${\bf w}_l^{\cdot \cdot} \in \mathbb R^d$ and ${\bf b}_l^\cdot \in \mathbb R^d$ are trainable weight and bias vectors.
		This is called the \textit{compress} operation, since the weight vectors project the cross feature matrix from $\mathbb R^{d \times d}$ space back to the feature spaces $\mathbb R^d$.
		Note that in Eq. (\ref{eq:compress}), the cross feature matrix is compressed along both horizontal and vertical directions (by operating on ${\bf C}_l$ and ${\bf C}_l^\top$) for the sake of symmetry, but we will provide more insights of the design in Section \ref{sec:unified_view}.
		For simplicity, the cross$\&$compress unit is denoted as:
		\begin{equation}
			[{\bf v}_{l+1}, {\bf e}_{l+1}] = \mathcal C ({\bf v}_l, {\bf e}_l),
		\end{equation}
		and we use a suffix $[{\bf v}]$ or $[{\bf e}]$ to distinguish its two outputs in the following of this paper.
		Through cross$\&$compress units, MKR can adaptively adjust the weights of knowledge transfer and learn the relevance between the two tasks.
		
		It should be noted that cross$\&$compress units should only exist in low-level layers of MKR, as shown in Figure \ref{fig:framework}.
		This is because:
		(1) In deep architectures, features usually transform from general to specific along the network, and feature transferability drops significantly in higher layers with increasing task dissimilarity \cite{yosinski2014transferable}.
		Therefore, sharing high-level layers risks to possible negative transfer, especially for the heterogeneous tasks in MKR.
		(2) In high-level layers of MKR, item features are mixed with user features, and entity features are mixed with relation features.
		The mixed features are not suitable for sharing since they have no explicit association.

	\subsection{Recommendation Module}
	\label{section:rm}
		The input of the recommendation module in MKR consists of two raw feature vectors $\bf u$ and $\bf v$ that describe user $u$ and item $v$, respectively.
		$\bf u$ and $\bf v$ can be customized as one-hot ID \cite{he2017neural}, attributes \cite{wang2018shine}, bag-of-words \cite{wang2015collaborative}, or their combinations, based on the application scenario.
		Given user $u$'s raw feature vector $\bf u$, we use an $L$-layer MLP to extract his latent condensed feature\footnote{We use the exponent notation $L$ in Eq. (\ref{eq:mlp}) and following equations in the rest of this paper for simplicity, but note that the parameters of $L$ layers are actually different.}:
		\begin{equation}
		\label{eq:mlp}
			{\bf u}_L = \mathcal M (\mathcal M (\cdots \mathcal M ({\bf u}))) = \mathcal M^L ({\bf u}),
		\end{equation}				
		where $\mathcal M ({\bf x}) = \sigma ({\bf W} {\bf x} + {\bf b} )$ is a fully-connected neural network layer\footnote{Exploring a more elaborate design of layers in the recommendation module is an important direction of future work.} with weight $\bf W$, bias $\bf b$, and nonlinear activation function $\sigma(\cdot)$.
		For item $v$, we use $L$ cross$\&$compress units to extract its feature:
		\begin{equation}
			{\bf v}_L = \mathbb E_{e \sim \mathcal S(v)} \left[ \mathcal C^L ({\bf v}, {\bf e})[{\bf v}] \right],
		\end{equation}
		where $\mathcal S(v)$ is the set of associated entities of item $v$.
		
		After having user $u$'s latent feature ${\bf u}_L$ and item $v$'s latent feature ${\bf v}_L$, we combine the two pathways by a predicting function $f_{RS}$, for example, inner product or an $H$-layer MLP.
		The final predicted probability of user $u$ engaging item $v$ is:
		\begin{equation}
		\hat y_{uv} = \sigma \big( f_{RS}({\bf u}_L, {\bf v}_L) \big).		\end{equation}

	\subsection{Knowledge Graph Embedding Module}
	\label{section:kgem}
		Knowledge graph embedding is to embed entities and relations into continuous vector spaces while preserving their structure.
		Recently, researchers have proposed a great many KGE methods, including translational distance models \cite{bordes2013translating,lin2015learning} and semantic matching models \cite{nickel2016holographic,liu2017analogical}.
		In MKR, we propose a deep semantic matching architecture for KGE module.
		Similar to the recommendation module, for a given knowledge triple $(h, r, t)$, we first utilize multiple cross$\&$compress units and nonlinear layers to process the raw feature vectors of head $h$ and relation $r$ (including ID \cite{lin2015learning}, types \cite{xie2016representation}, textual description \cite{wang2014knowledge}, etc.), respectively.
		Their latent features are then concatenated together, followed by a $K$-layer MLP for predicting tail $t$:
		\begin{equation}
		\begin{split}
			&{\bf h}_L = \mathbb E_{v \sim \mathcal S(h)} \left[ \mathcal C^L ({\bf v}, {\bf h})[{\bf e}] \right],\\
			&{\bf r}_L = \mathcal M^L ({\bf r}),\\
			&{\bf \hat t} = \mathcal M^K \left(
			\begin{bmatrix}
				{\bf h}_L\\
				{\bf r}_L
			\end{bmatrix} \right),
		\end{split}
		\end{equation}
		where $\mathcal S(h)$ is the set of associated items of entity $h$, and $\bf \hat t$ is the predicted vector of tail $t$.
		Finally, the score of the triple $(h, r, t)$ is calculated using a score (similarity) function $f_{KG}$:
		\begin{equation}
			score(h, r, t) = f_{KG}({\bf t}, {\bf \hat t}),
		\end{equation}
		where $\bf t$ is the real feature vector of $t$.
		In this paper, we use the normalized inner product $f_{KG}({\bf t}, {\bf \hat t}) = \sigma({\bf t}^\top {\bf \hat t})$ as the choice of score function \cite{misra2016cross}, but other forms of (dis)similarity metrics can also be applied here such as Kullback–Leibler divergence.

	\subsection{Learning Algorithm}
		The complete loss function of MKR is as follows:
		\begin{equation}
		\label{eq:loss}
		\begin{split}
			\mathcal L =& \mathcal L_{RS} + \mathcal L_{KG} + \mathcal L_{REG}\\
			=& \sum_{u \in \mathcal U, v \in \mathcal V} \mathcal J(\hat y_{uv}, y_{uv})\\
			&- \lambda_1 \Big( \sum_{(h, r, t) \in \mathcal G} score(h, r, t) - \sum_{(h', r, t') \notin \mathcal G} score(h', r, t') \Big)\\
			&+ \lambda_2 \| {\bf W} \|_2^2.
		\end{split}
		\end{equation}
			
		In Eq. (\ref{eq:loss}), the first term measures loss in the recommendation module, where $u$ and $v$ traverse the set of users and the items, respectively, and $\mathcal J$ is the cross-entropy function.
		The second term calculates the loss in the KGE module, in which we aim to increase the score for all true triples while reducing the score for all false triples.
		The last item is the regularization term for preventing over-fitting, $\lambda_1$ and $\lambda_2$ are the balancing parameters.\footnote{$\lambda_1$ can be seen as the ratio of two learning rates for the two tasks.}
			
		Note that the loss function in Eq. (\ref{eq:loss}) traverses all possible user-item pairs and knowledge triples.
		To make computation more efficient, following \cite{mikolov2013distributed}, we use a negative sampling strategy during training.
		\begin{algorithm}[t]
  		\caption{Multi-Task Training for MKR}
			\begin{algorithmic}[1]
				\REQUIRE{Interaction matrix $\bf Y$, knowledge graph $\mathcal G$}
				\ENSURE{Prediction function $\mathcal F(u, v | \Theta, \bf Y, \mathcal G)$}
				\STATE Initialize all parameters
				\FOR{number of training iteration}
					\STATEx \quad // \textit{recommendation task}
					\FOR{$t$ steps}
						\STATE Sample minibatch of positive and negative interactions from $\bf Y$;
						\STATE Sample $e \sim \mathcal S(v)$ for each item $v$ in the minibatch;
						\STATE Update parameters of $\mathcal F$ by gradient descent on Eq. (1)-(6), (9);
					\ENDFOR
					\STATEx \quad // \textit{knowledge graph embedding task}
					\STATE Sample minibatch of true and false triples from $\mathcal G$;
					\STATE Sample $v \sim \mathcal S(h)$ for each head $h$ in the minibatch;
					\STATE Update parameters of $\mathcal F$ by gradient descent on Eq. (1)-(3), (7)-(9);
				\ENDFOR
			\end{algorithmic}
		\end{algorithm}		
		The learning algorithm of MKR is presented in Algorithm 1, in which a training epoch consists of two stages: recommendation task (line 3-7) and KGE task (line 8-10).
		In each iteration, we repeat training on recommendation task for $t$ times ($t$ is a hyper-parameter and normally $t > 1$) before training on KGE task once in each epoch, since we are more focused on improving recommendation performance.
		We will discuss the choice of $t$ in the experiments section.

\section{Theoretical Analysis}
	In this section, we prove that cross$\&$compress units have sufficient capability of polynomial approximation.
	We also show that MKR is a generalized framework over several representative methods of recommender systems and multi-task learning.
	
	\subsection{Polynomial Approximation}
		According to the Weierstrass approximation theorem \cite{rudin1964principles}, any function under certain smoothness assumption can be approximated by a polynomial to an arbitrary accuracy.
		Therefore, we examine the ability of high-order interaction approximation of the cross$\&$compress unit. 
		We show that cross$\&$compress units can model the order of item-entity feature interaction up to exponential degree:
		
		\begin{theorem}
		\label{thm:1}
			Denote the input of item and entity in MKR network as ${\bf v} = [v_1 \cdots \ v_d]^\top$ and ${\bf e} = [e_1 \ \cdots \ e_d]^\top$, respectively.
			Then the cross terms about ${\bf v}$ and ${\bf e}$ in $\|{\bf v}_L\|_1$ and $\|{\bf e}_L\|_1$ (the L1-norm of ${\bf v}_L$ and ${\bf e}_L$) with maximal degree is $k_{\bm \alpha, \bm \beta} v_1^{\alpha_1} \cdots v_d^{\alpha_d} e_1^{\beta_1} \cdots e_d^{\beta_d}$, where $k_{\bm \alpha, \bm \beta} \in \mathbb R$, $\alpha_i, \beta_i \in \mathbb N$ for $i \in \{1, \cdots, d\}$, $\alpha_1 + \cdots + \alpha_d = 2^{L-1}$, and $\beta_1 + \cdots + \beta_d = 2^{L-1}$ ($L \geq 1, {\bf v}_0 = {\bf v}, {\bf e}_0 = {\bf e}$).
		\end{theorem}
		
		In recommender systems, $\prod_{i=1}^d v_i^{\alpha_i} e_i^{\beta_i}$ is also called \textit{combinatorial} feature, as it measures the interactions of multiple original features.
		Theorem \ref{thm:1} states that cross$\&$compress units can automatically model the combinatorial features of items and entities for sufficiently high order, which demonstrates the superior approximation capacity of MKR as compared with existing work such as Wide$\&$Deep \cite{cheng2016wide}, factorization machines \cite{rendle2010factorization, rendle2012factorization} and DCN \cite{wang2017deep}.
		The proof of Theorem \ref{thm:1} is provided in the Appendix.
		Note that Theorem \ref{thm:1} gives a theoretical view of the polynomial approximation ability of the cross$\&$compress unit rather than providing guarantees on its actual performance.
		We will empirically evaluate the cross$\&$compress unit in the experiments section.

	\subsection{Unified View of Representative Methods}
	\label{sec:unified_view}
		In the following we provide a unified view of several representative models in recommender systems and multi-task learning, by showing that they are restricted versions of or theoretically related to MKR.
		This justifies the design of cross$\&$compress unit and conceptually explains its strong empirical performance as compared to baselines.
		
		\subsubsection{Factorization machines}
		Factorization machines \cite{rendle2010factorization, rendle2012factorization} are a generic method for recommender systems.
		Given an input feature vector, FMs model all interactions between variables in the input vector using factorized parameters, thus being able to estimate interactions in problems with huge sparsity such as recommender systems.
		The model equation for a 2-degree factorization machine is defined as
		\begin{equation}		
			\hat y({\bf x}) = w_0 + \sum\nolimits_{i=1}^d w_i x_i + \sum\nolimits_{i=1}^d \sum\nolimits_{j=i+1}^d \langle {\bf v}_i, {\bf v}_j \rangle x_i x_j,
		\end{equation}
		where $x_i$ is the $i$-th unit of input vector $\bf x$, $w_\cdot$ is weight scalar, ${\bf v}_\cdot$ is weight vector, and $\langle \cdot, \cdot \rangle$ is dot product of two vectors.
		We show that the essence of FM is conceptually similar to an 1-layer cross$\&$compress unit:
		
		\begin{proposition}
		\label{prop:1}
			The L1-norm of ${\bf v}_1$ and ${\bf e}_1$ can be written as the following form:
			\begin{equation}
				\|{\bf v}_1\|_1 \left( or \ \|{\bf e}_1\|_1 \right) = \left| b + \sum\nolimits_{i=1}^d \sum\nolimits_{j=1}^d \langle w_i, w_j \rangle v_i e_j \right|,
			\end{equation}
			where $\langle w_i, w_j \rangle = w_i + w_j$ is the sum of two scalars.
		\end{proposition}
		
		It is interesting to notice that, instead of factorizing the weight parameter of $x_i x_j$ into the dot product of two vectors as in FM, the weight of term $v_i e_j$ is factorized into the sum of two scalars in cross$\&$compress unit to reduce the number of parameters and increase robustness of the model.

		\subsubsection{Deep$\&$Cross Network}
		DCN \cite{wang2017deep} learns explicit and high-order cross features by introducing the layers:
		\begin{equation}
			{\bf x}_{l+1} = {\bf x}_0 {\bf x}_l^\top {\bf w}_l + {\bf x}_l + {\bf b}_l,
		\end{equation}			
		where ${\bf x}_l$, ${\bf w}_l$, and ${\bf b}_l$ are representation, weight, and bias of the $l$-th layer.
		We demonstrate the link between DCN and MKR by the following proposition:
		
		\begin{proposition}
		\label{prop:2}
			In the formula of ${\bf v}_{l+1}$ in Eq. (\ref{eq:compress}), if we restrict ${\bf w}_l^{VV}$ in the first term to satisfy ${\bf e}_l^\top {\bf w}_l^{VV} = 1$ and restrict ${\bf e}_l$ in the second term to be ${\bf e}_0$ (and impose similar restrictions on ${\bf e}_{l+1}$), the cross$\&$compress unit is then conceptually equivalent to DCN layer in the sense of multi-task learning:
			\begin{equation}
			\label{eq:dcn}
			\begin{split}
				&{\bf v}_{l+1} = {\bf e}_0 {\bf v}_l^\top {\bf w}_l^{EV} + {\bf v}_l + {\bf b}_l^V,\\
				&{\bf e}_{l+1} = {\bf v}_0 {\bf e}_l^\top {\bf w}_l^{VE} + {\bf e}_l + {\bf b}_l^E.
			\end{split}
			\end{equation}
		\end{proposition}
		
		It can be proven that the polynomial approximation ability of the above DCN-equivalent version (i.e., the maximal degree of cross terms in ${\bf v}_l$ and ${\bf e}_l$) is $O(l)$, which is weaker than original cross$\&$compress units with $O(2^l)$ approximation ability.
		
		\subsubsection{Cross-stitch Networks}
		Cross-stitch networks \cite{misra2016cross} is a multi-task learning model in convolutional networks, in which the designed cross-stitch unit can learn a combination of shared and task-specific representations between two tasks.
		Specifically, given two activation maps $x_A$ and $x_B$ from layer $l$ for both the tasks, cross-stitch networks learn linear combinations $\tilde x_A$ and $\tilde x_B$ of both the input activations and feed these combinations as input to the next layers' filters.
		The formula at location $(i, j)$ in the activation map is
		
		\begin{equation}
		\label{eq:csn}
			\begin{bmatrix}
   				\tilde x_A^{ij}\\[5pt]
   				\tilde x_B^{ij}
  			\end{bmatrix}
  			=
  			\begin{bmatrix}
   				\alpha_{AA} & \alpha_{AB}\\
   				\alpha_{BA} & \alpha_{BB}
  			\end{bmatrix}
  			\begin{bmatrix}
   				x_A^{ij}\\[5pt]
   				x_B^{ij}
  			\end{bmatrix},
		\end{equation}
		where $\alpha$'s are trainable transfer weights of representations between task A and task B.
		We show that the cross-stitch unit in Eq. (\ref{eq:csn}) is a simplified version of our cross$\&$compress unit by the following proposition:
		
		\begin{proposition}
		\label{prop:3}
			If we omit all biases in Eq. (\ref{eq:compress}), the cross$\&$compress unit can be written as
			\begin{equation}
			\label{eq:cross-stitch}
				\begin{bmatrix}
   				{\bf v}_{l+1}\\
   				{\bf e}_{l+1}
  			\end{bmatrix}
  			=
  			\begin{bmatrix}
   				{\bf e}_l^\top {\bf w}_l^{VV} & {\bf v}_l^\top {\bf w}_l^{EV}\\[5pt]
   				{\bf e}_l^\top {\bf w}_l^{VE} & {\bf v}_l^\top {\bf w}_l^{EE}
  			\end{bmatrix}
  			\begin{bmatrix}
   				{\bf v}_l\\
   				{\bf e}_l
  			\end{bmatrix}.
			\end{equation}
		\end{proposition}
		
		The transfer matrix in Eq. (\ref{eq:cross-stitch}) serves as the cross-stitch unit $[\alpha_{AA} \ \alpha_{AB}; \ \alpha_{BA} \ \alpha_{BB}]$ in Eq. (\ref{eq:csn}).
		Like cross-stitch networks, MKR network can decide to make certain layers task specific by setting ${\bf v}_l^\top {\bf w}_l^{EV}$ ($\alpha_{AB}$) or ${\bf e}_l^\top {\bf w}_l^{VE}$ ($\alpha_{BA}$) to zero, or choose a more shared representation by assigning a higher value to them.
		But the transfer matrix is more fine-grained in cross$\&$compress unit, because the transfer weights are replaced from scalars to dot products of two vectors.
		It is rather interesting to notice that Eq. (\ref{eq:cross-stitch}) can also be regarded as an \textit{attention mechanism} \cite{bahdanau2014neural}, as the computation of transfer weights involves the feature vectors ${\bf v}_l$ and ${\bf e}_l$ themselves.

\section{Experiments}
	In this section, we evaluate the performance of MKR in four real-world recommendation scenarios: movie, book, music, and news\footnote{The source code is available at \url{https://github.com/hwwang55/MKR}.}.
	
	\begin{table*}[t]
		\setlength{\tabcolsep}{8pt}
		\centering
		\caption{Basic statistics and hyper-parameter settings for the four datasets.}
		\begin{tabular}{c|cccc|c}
			\hline
			Dataset & \# users & \# items & \# interactions & \# KG triples & Hyper-parameters\\
			\hline
			MovieLens-1M & 6,036 & 2,347 & 753,772 & 20,195 & $L = 1$, $d = 8$, $t = 3$, $\lambda_1 = 0.5$\\
			Book-Crossing & 17,860 & 14,910 & 139,746 & 19,793 & $L = 1$, $d = 8$, $t = 2$, $\lambda_1 = 0.1$\\
			Last.FM & 1,872 & 3,846 & 42,346 & 15,518 & $L$ = 2, $d = 4$, $t = 2$, $\lambda_1 = 0.1$\\
			Bing-News & 141,487 & 535,145 & 1,025,192 & 1,545,217 & $L = 3$, $d = 16$, $t = 5$, $\lambda_1 = 0.2$\\
			\hline
		\end{tabular}
		\label{table:statistics}
	\end{table*}	
		
	\subsection{Datasets}
		We utilize the following four datasets in our experiments:
		\begin{itemize}
			\item
				\textbf{MovieLens-1M}\footnote{\url{https://grouplens.org/datasets/movielens/1m/}} is a widely used benchmark dataset in movie recommendations, which consists of approximately 1 million explicit ratings (ranging from 1 to 5) on the MovieLens website.
			\item
				\textbf{Book-Crossing}\footnote{\url{http://www2.informatik.uni-freiburg.de/~cziegler/ BX/}} dataset contains 1,149,780 explicit ratings (ranging from 0 to 10) of books in the Book-Crossing community.
			\item
				\textbf{Last.FM}\footnote{\url{https://grouplens.org/datasets/hetrec-2011/}} dataset contains musician listening information from a set of 2 thousand users from Last.fm online music system.
			\item
				\textbf{Bing-News} dataset contains 1,025,192 pieces of implicit feedback collected from the server logs of Bing News\footnote{\url{https://www.bing.com/news}} from October 16, 2016 to August 11, 2017.
				Each piece of news has a title and a snippet.
		\end{itemize}
		
		Since MovieLens-1M, Book-Crossing, and Last.FM are explicit feedback data (Last.FM
provides the listening count as weight for each user-item interaction), we transform them into implicit feedback where each entry is marked with 1 indicating that the user has rated the item positively, and sample an unwatched set marked as 0 for each user.
		The threshold of positive rating is 4 for MovieLens-1M, while no threshold is set for Book-Crossing and Last.FM due to their sparsity.

		We use Microsoft Satori to construct the KG for each dataset.
		We first select a subset of triples from the whole KG with a confidence level greater than 0.9.
		For MovieLens-1M and Book-Crossing, we additionally select a subset of triples from the sub-KG whose relation name contains "film" or "book" respectively to further reduce KG size.
		
		Given the sub-KGs, for MovieLens-1M, Book-Crossing, and Last.FM, we collect IDs of all valid movies, books, or musicians by matching their names with tail of triples (\textit{head, film.film.name, tail}), \textit{(head, book.book.title, tail)}, or \textit{(head, type.object.name, tail)}, respectively.
		For simplicity, items with no matched or multiple matched entities are excluded.
		We then match the IDs with the head and tail of all KG triples and select all well-matched triples from the sub-KG.
		The constructing process is similar for Bing-News except that: (1) we use entity linking tools to extract entities in news titles; (2) we do not impose restrictions on the names of relations since the entities in news titles are not within one particular domain.
		The basic statistics of the four datasets are presented in Table \ref{table:statistics}.
		Note that the number of users, items, and interactions are smaller than original datasets since we filtered out items with no corresponding entity in the KG.

	\subsection{Baselines}
		We compare our proposed MKR with the following baselines.
		Unless otherwise specified, the hyper-parameter settings of baselines are the same as reported in their original papers or as default in their codes.
		
		\begin{itemize}
			\item
				\textbf{PER} \cite{yu2014personalized} treats the KG as heterogeneous information networks and extracts meta-path based features to represent the connectivity between users and items.
				In this paper, we use manually designed user-item-attribute-item paths as
features, i.e., "user-movie-director-movie", "user-movie-genre-movie", and "user-movie-star-movie" for MovieLens-20M; "user-book-author-book" and "user-book-genre-book" for Book-Crossing; "user-musician-genre-musician", "user-musician-country-musician", and "user-musician-age-musician" (age is discretized) for Last.FM.
				Note that PER cannot be applied to news recommendation because it's hard to pre-define meta-paths for entities in news.
			\item
				\textbf{CKE} \cite{zhang2016collaborative} combines CF with structural, textual, and visual knowledge in a unified framework for recommendation.
				We implement CKE as CF plus structural knowledge module in this paper.
				The dimension of user and item embeddings for the four datasets are set as 64, 128, 32, 64, respectively.
				The dimension of entity embeddings is $32$.
			\item
				\textbf{DKN} \cite{wang2018dkn} treats entity embedding and word embedding as multiple channels and combines them together in CNN for CTR prediction.
				In this paper, we use movie/book names and news titles as textual input for DKN.
				The dimension of word embedding and entity embedding is 64, and the number of filters is 128 for each window size 1, 2, 3.
			\item				
				\textbf{RippleNet} \cite{wang2018ripple} is a memory-network-like approach that propagates users’ preferences on the knowledge graph for recommendation.
				The hyper-parameter settings for Last.FM are $d=8$, $H=2$, $\lambda_1 = 10^{-6}$, $\lambda_2=0.01$, $\eta=0.02$.
			\item
				\textbf{LibFM} \cite{rendle2012factorization} is a widely used feature-based factorization model.
				We concatenate the raw features of users and items as well as the corresponding averaged entity embeddings learned from TransR \cite{lin2015learning} as input for LibFM.
				The dimension is \{1, 1, 8\} and the number of training epochs is 50.
				The dimension of TransR is 32.
			\item
				\textbf{Wide$\&$Deep} \cite{cheng2016wide} is a deep recommendation model combining a (wide) linear channel with a (deep) nonlinear channel.
				The input for Wide$\&$Deep is the same as in LibFM.
				The dimension of user, item, and entity is 64, and we use a two-layer deep channel with dimension of 100 and 50 as well as a wide channel.
		\end{itemize}
		
		\begin{figure}[t]
			\centering
            \begin{subfigure}[b]{0.23\textwidth}
                \includegraphics[width=\textwidth]{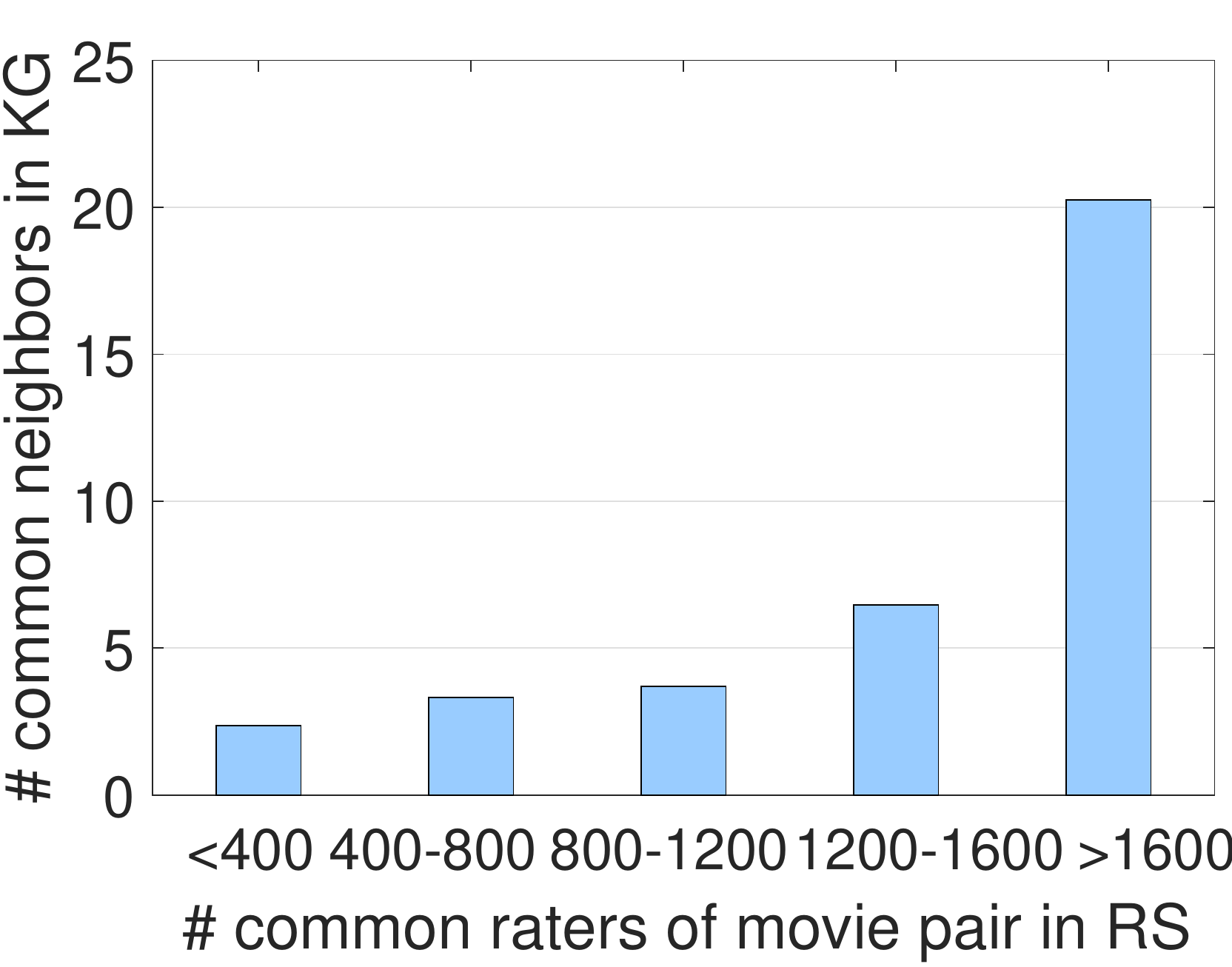}
                \caption{RS to KG}
                \label{fig:case_study_1}
            \end{subfigure}
            \hfill
            \begin{subfigure}[b]{0.23\textwidth}
                \includegraphics[width=\textwidth]{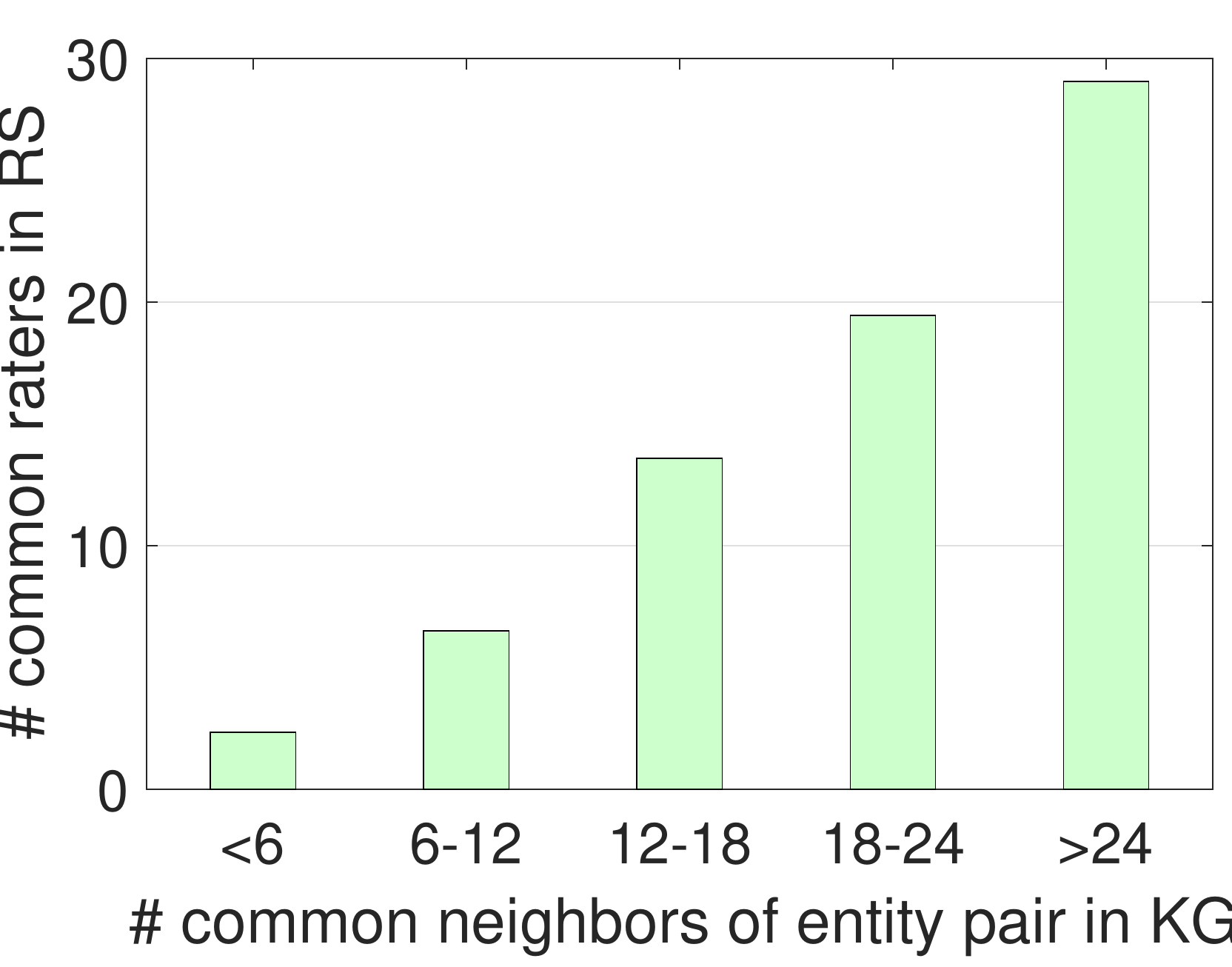}
                \caption{KG to RS}
                \label{fig:case_study_2}
            \end{subfigure}
            \caption{The correlation between the number of common neighbors of an item pair in KG and their number of common raters in RS.}
        \end{figure}
        
        \begin{table*}[t]
			\setlength{\tabcolsep}{4pt}
            \centering
            \caption{The results of $AUC$ and $Accuracy$ in CTR prediction.}
            \begin{tabular}{c|cccccccc}
            	\hline
            	\multirow{2}{*}{Model} & \multicolumn{2}{c}{MovieLens-1M} & \multicolumn{2}{c}{Book-Crossing} & \multicolumn{2}{c}{Last.FM} & \multicolumn{2}{c}{Bing-News} \\
            	\cline{2-9}
            	& $AUC$ & $ACC$ & $AUC$ & $ACC$ & $AUC$ & $ACC$ & $AUC$ & $ACC$ \\
            	\hline
            	PER & 0.710 (-22.6\%) & 0.664 (-21.2\%) & 0.623 (-15.1\%) & 0.588 (-16.7\%) & 0.633 (-20.6\%) & 0.596 (-20.7\%) & - & - \\
            	CKE & 0.801 (-12.6\%) & 0.742 (-12.0\%) & 0.671 (-8.6\%) & 0.633 (-10.3\%) & 0.744 (-6.6\%) & 0.673  (-10.5\%) & 0.553 (-19.7\%) & 0.516 (-20.0\%) \\
            	DKN & 0.655 (-28.6\%)  & 0.589 (-30.1\%) & 0.622 (-15.3\%) & 0.598 (-15.3\%) & 0.602 (-24.5\%) & 0.581 (-22.7\%) & 0.667 (-3.2\%) & 0.610 (-5.4\%) \\
				RippleNet & \textbf{0.920} (+0.3\%) & 0.842 (-0.1\%) & 0.729 (-0.7\%) & 0.662 (-6.2\%) & 0.768 (-3.6\%) & 0.691 (-8.1\%) & 0.678 (-1.6\%) & 0.630 (-2.3\%)\\            	
            	LibFM & 0.892 (-2.7\%) & 0.812 (-3.7\%) & 0.685 (-6.7\%) & 0.640 (-9.3\%) & 0.777 (-2.5\%) & 0.709 (-5.7\%) & 0.640 (-7.1\%) & 0.591 (-8.4\%) \\
            	Wide$\&$Deep & 0.898 (-2.1\%) & 0.820 (-2.7\%) & 0.712 (-3.0\%) & 0.624 (-11.6\%) & 0.756 (-5.1\%) & 0.688 (-8.5\%) & 0.651 (-5.5\%) & 0.597 (-7.4\%) \\
            	\hline
            	MKR & 0.917 & \textbf{0.843} & \textbf{0.734} & \textbf{0.704} & \textbf{0.797} & \textbf{0.752} & \textbf{0.689} & \textbf{0.645} \\
            	MKR-1L & - & - & - & - & 0.795 (-0.3\%) & 0.749 (-0.4\%) & 0.680 (-1.3\%) & 0.631 (-2.2\%) \\
            	MKR-DCN & 0.883 (-3.7\%) & 0.802 (-4.9\%) & 0.705 (-4.3\%) & 0.676 (-4.2\%) & 0.778 (-2.4\%) & 0.730 (-2.9\%) & 0.671 (-2.6\%) & 0.614 (-4.8\%) \\
            	MKR-stitch & 0.905 (-1.3\%) & 0.830 (-1.5\%) & 0.721 (-2.2\%) & 0.682 (-3.4\%) & 0.772 (-3.1\%) & 0.725  (-3.6\%) & 0.674 (-2.2\%) & 0.621 (-3.7\%) \\
            	\hline
			\end{tabular}
			\label{table:ctr}
		\end{table*}

	\subsection{Experiments setup}
		In MKR, we set the number of high-level layers $K = 1$, $f_{RS}$ as inner product, and $\lambda_2 = 10^{-6}$ for all three datasets, and other hyper-parameter are given in Table \ref{table:statistics}.
		The settings of hyper-parameters are determined by optimizing $AUC$ on a validation set.
		For each dataset, the ratio of training, validation, and test set is $6 : 2 : 2$.
		Each experiment is repeated $3$ times, and the average performance is reported.
		We evaluate our method in two experiment scenarios:
		(1) In click-through rate (CTR) prediction, we apply the trained model to each piece of interactions in the test set and output the predicted click probability.
		We use $AUC$ and $Accuracy$ to evaluate the performance of CTR prediction.
		(2) In top-$K$ recommendation, we use the trained model to select $K$ items with highest predicted click probability for each user in the test set, and choose $Precision@K$ and $Recall@K$ to evaluate the recommended sets.
	
		\begin{figure*}[t]
			\centering
			\begin{subfigure}[b]{0.95\textwidth}
                \includegraphics[width=\textwidth]{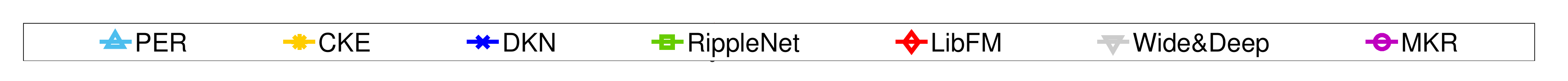}
                \vspace{-0.2in}
            \end{subfigure}
            \hfill
            \begin{subfigure}[b]{0.24\textwidth}
                \includegraphics[width=\textwidth]{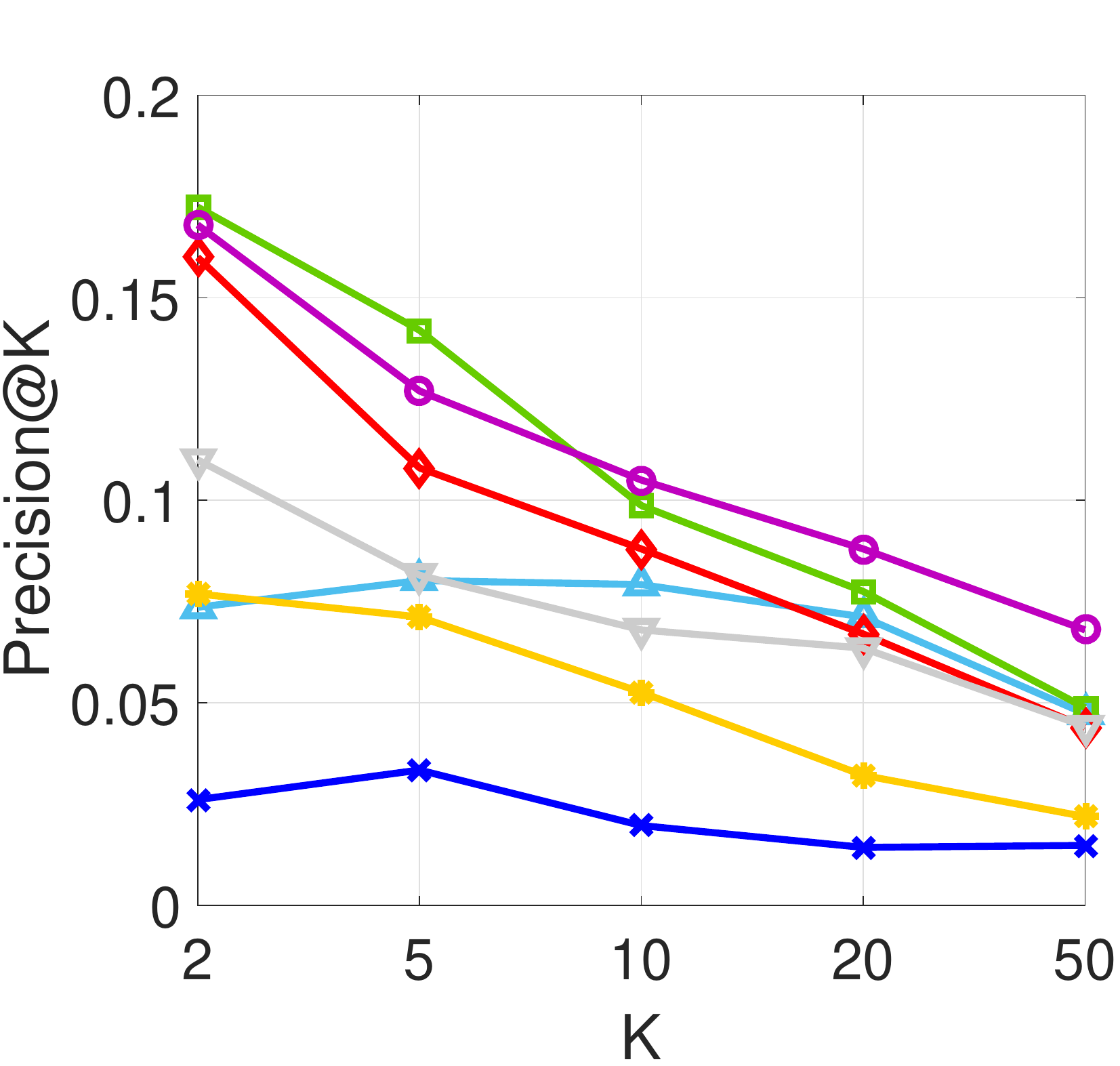}
                \caption{MovieLens-1M}
            \end{subfigure}
            \hfill
            \begin{subfigure}[b]{0.24\textwidth}
                \includegraphics[width=\textwidth]{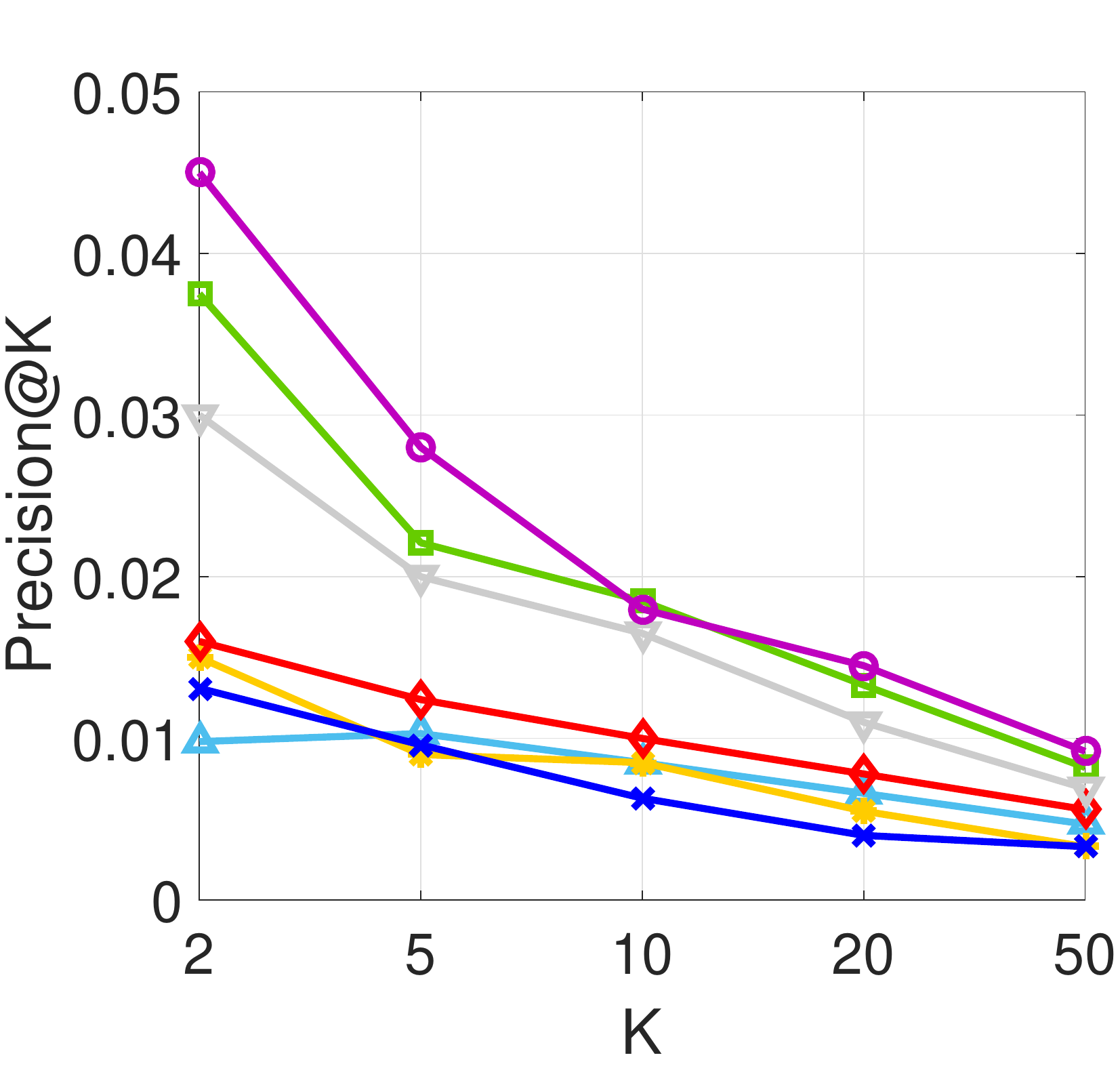}
                \caption{Book-Crossing}
            \end{subfigure}
            \hfill
            \begin{subfigure}[b]{0.24\textwidth}
                \includegraphics[width=\textwidth]{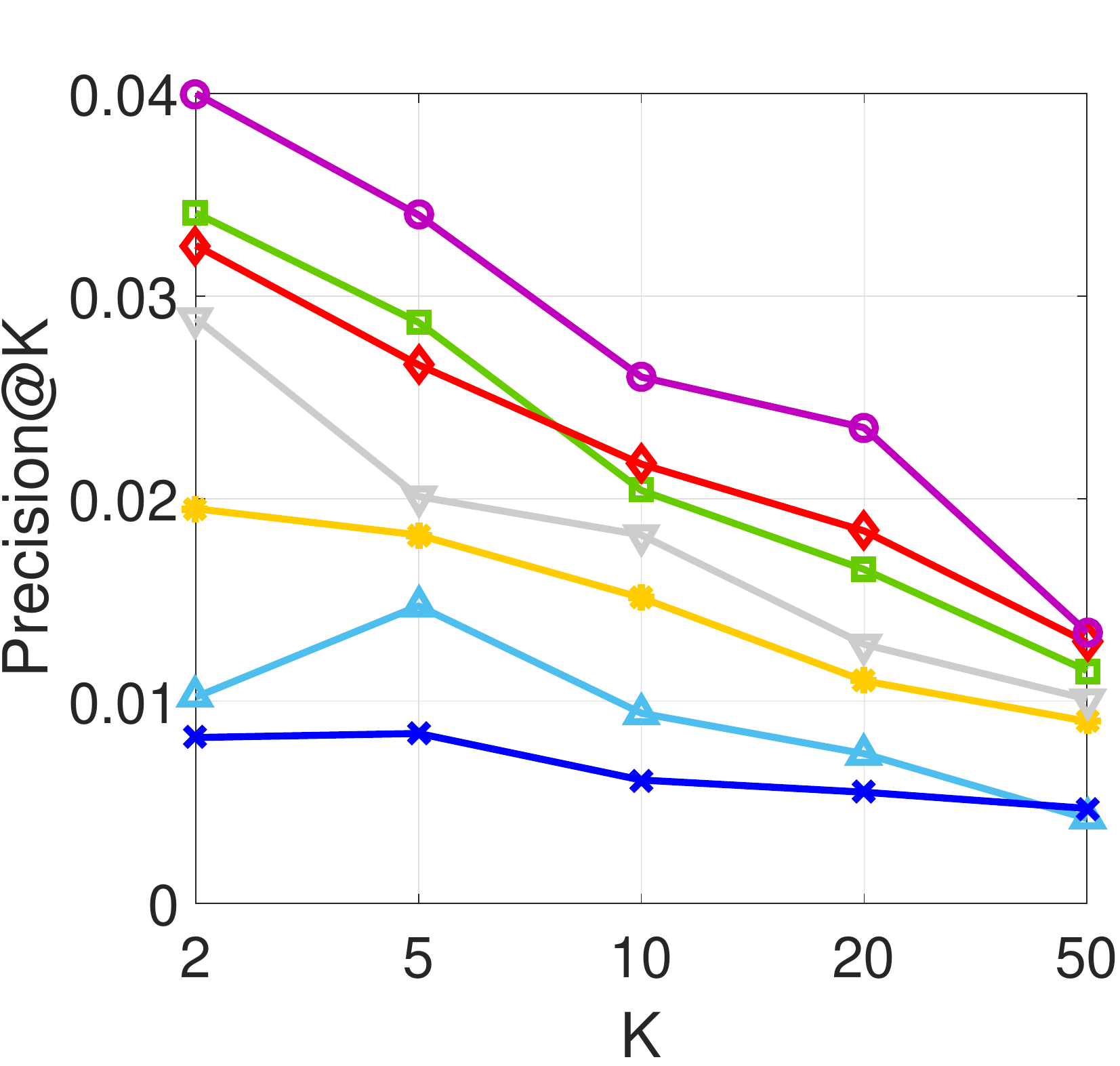}
                \caption{Last.FM}
            \end{subfigure}
            \hfill
            \begin{subfigure}[b]{0.24\textwidth}
                \includegraphics[width=\textwidth]{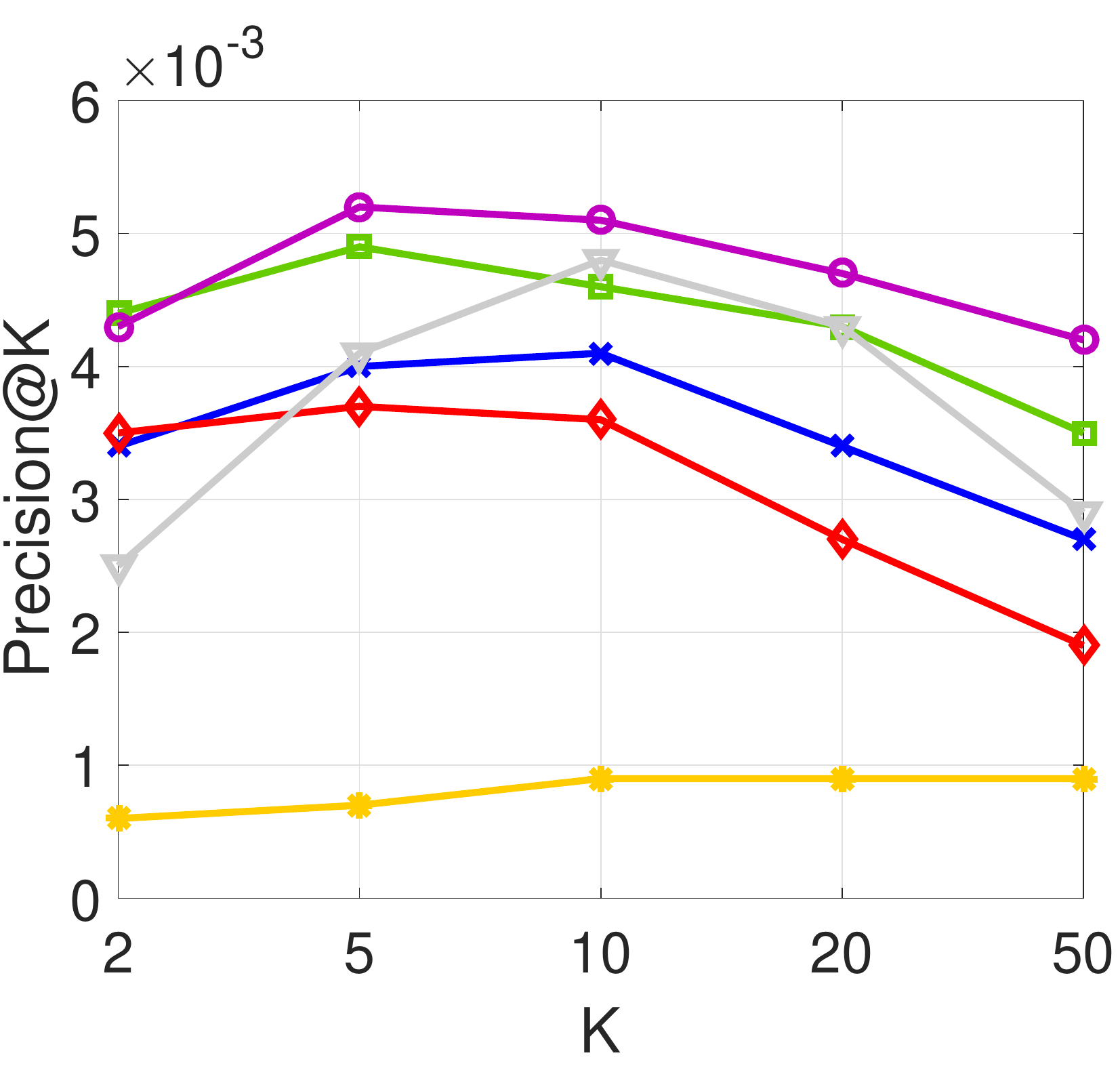}
                \caption{Bing-News}
            \end{subfigure}
            \caption{The results of $Precision@K$ in top-$K$ recommendation.}
            \label{fig:precision}
        \end{figure*}
        
        \begin{figure*}[t]
			\centering
			\begin{subfigure}[b]{0.95\textwidth}
                \includegraphics[width=\textwidth]{charts/legend.eps}
                \vspace{-0.15in}
            \end{subfigure}
            \hfill
            \begin{subfigure}[b]{0.24\textwidth}
                \includegraphics[width=\textwidth]{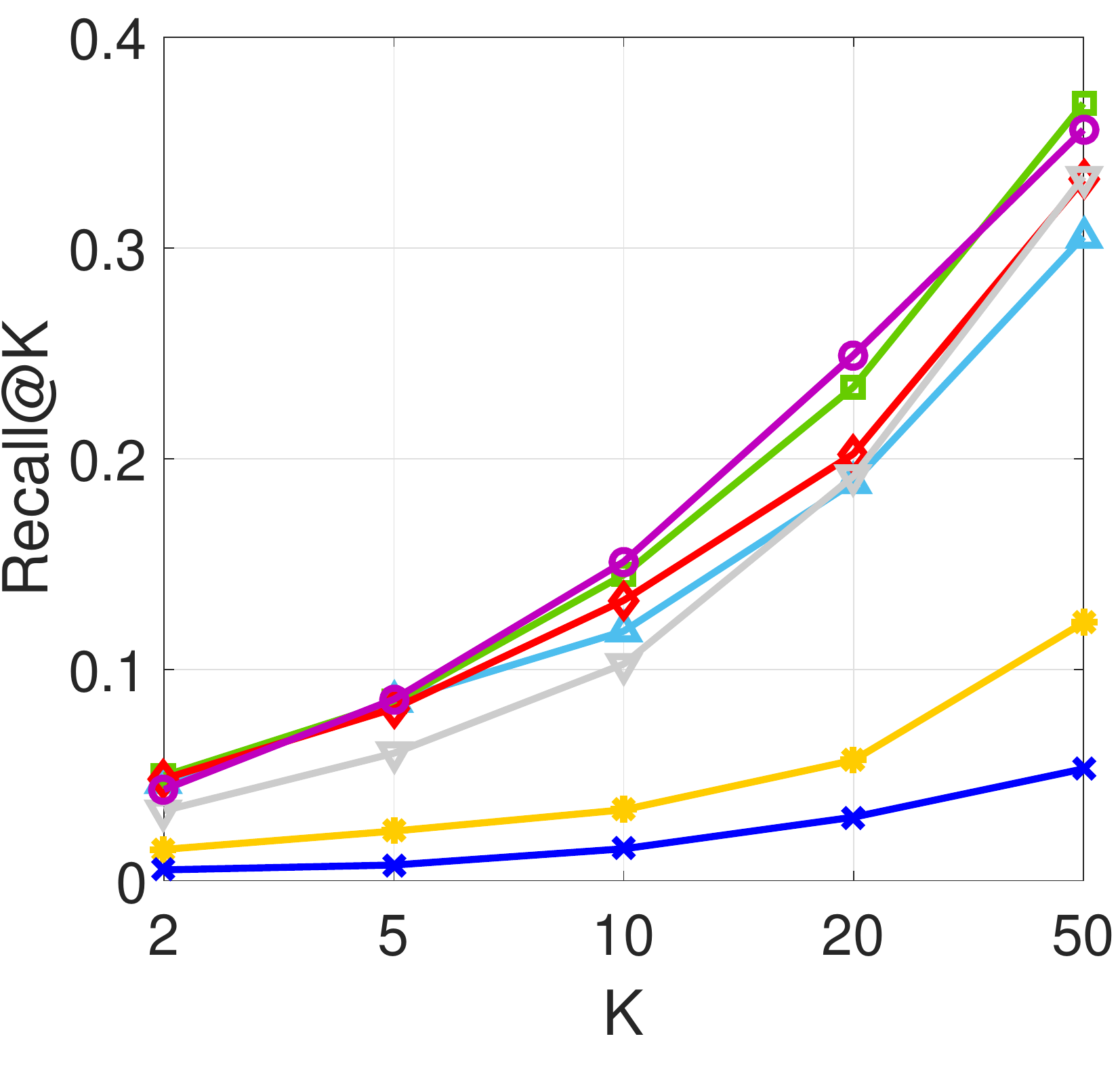}
                \caption{MovieLens-1M}
            \end{subfigure}
            \hfill
            \begin{subfigure}[b]{0.24\textwidth}
                \includegraphics[width=\textwidth]{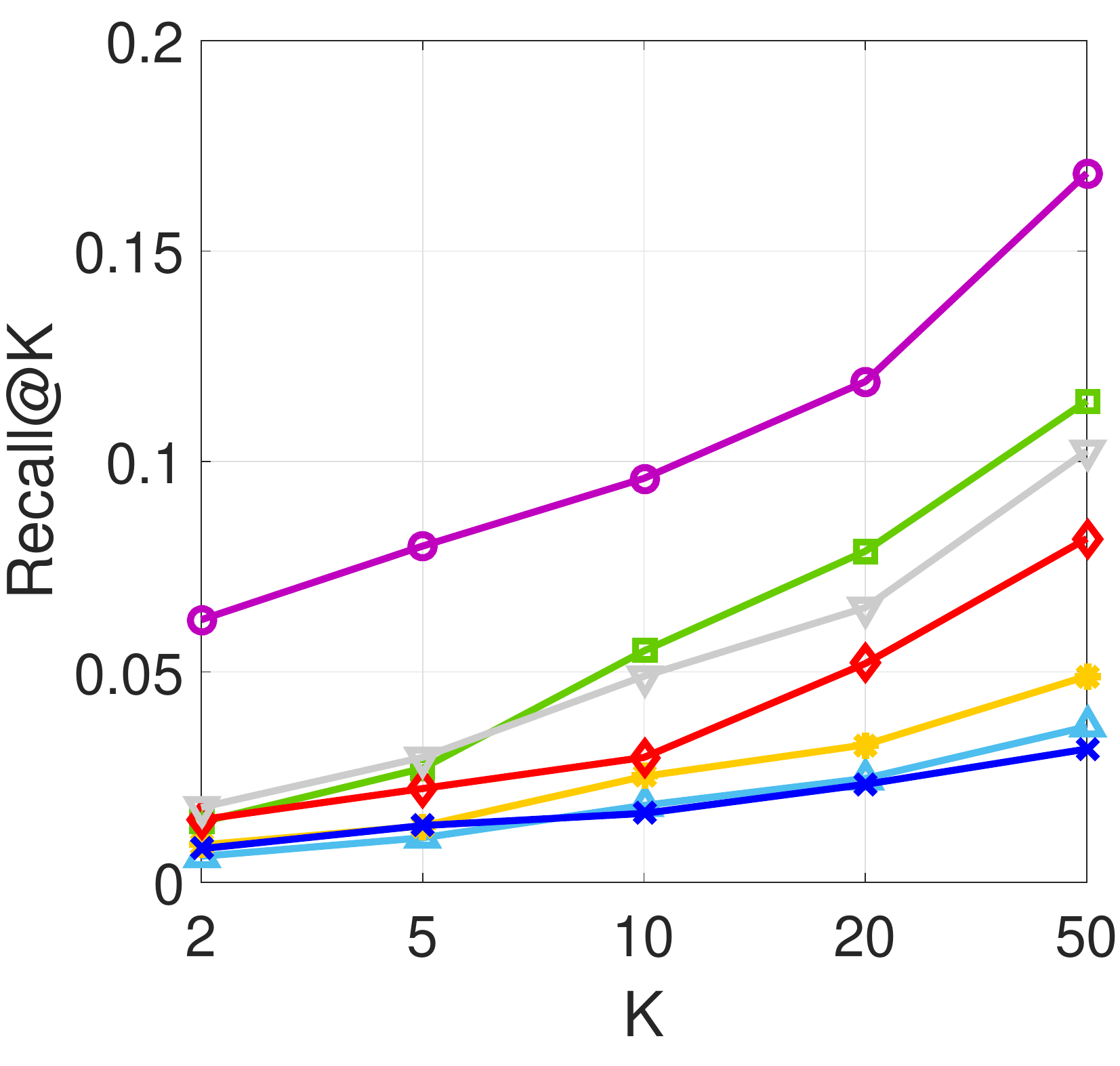}
                \caption{Book-Crossing}
            \end{subfigure}
            \hfill
            \begin{subfigure}[b]{0.24\textwidth}
                \includegraphics[width=\textwidth]{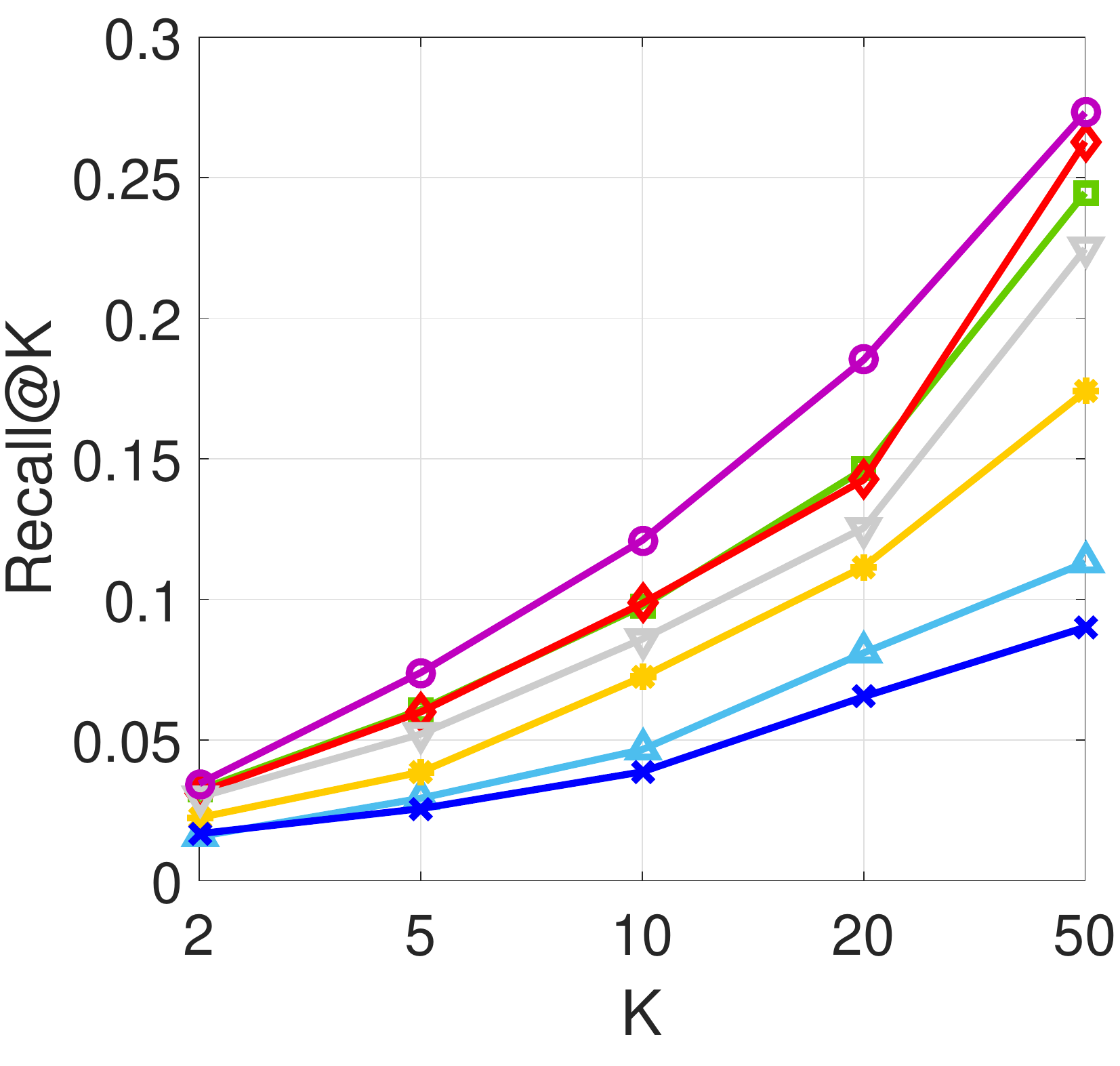}
                \caption{Last.FM}
            \end{subfigure}
            \hfill
            \begin{subfigure}[b]{0.24\textwidth}
                \includegraphics[width=\textwidth]{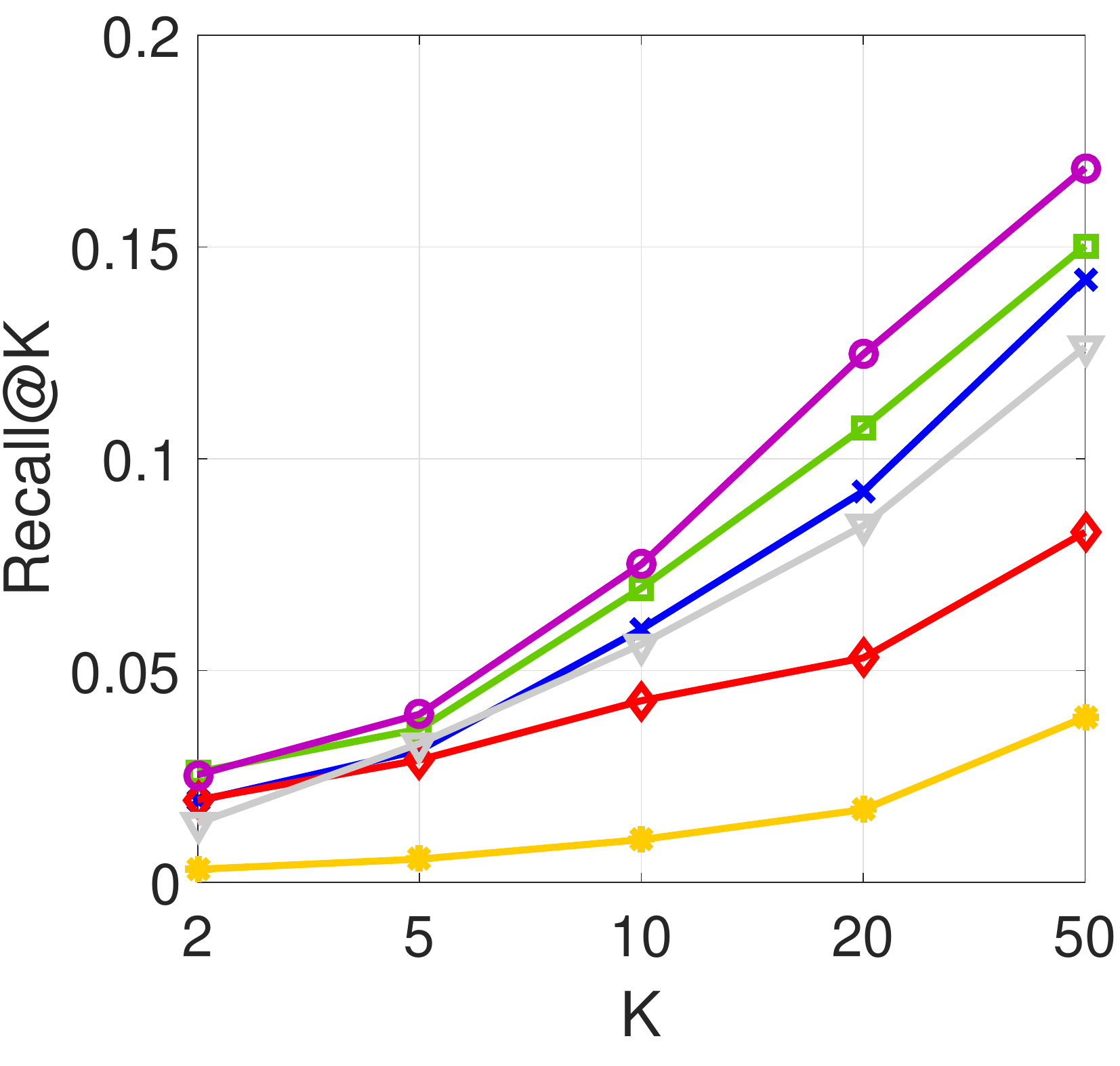}
                \caption{Bing-News}
            \end{subfigure}
            \caption{The results of $Recall@K$ in top-$K$ recommendation.}
            \label{fig:recall}
        \end{figure*}

	\subsection{Empirical study}        
		We conduct an empirical study to investigate the correlation of items in RS and their corresponding entities in KG.
		Specifically, we aim to reveal how the number of common neighbors of an item pair in KG changes with their number of common raters in RS.
		To this end, we first randomly sample 1 million item pairs from MovieLens-1M.
		We then classify each pair into 5 categories based on the number of their common raters in RS, and count their average number of common neighbors in KG for each category.
		The result is presented in Figure \ref{fig:case_study_1}, which clearly shows that \textit{if two items have more common raters in RS, they are likely to share more common neighbors in KG}.
		Figure \ref{fig:case_study_2} shows the positive correlation from an opposite direction.
		The above findings empirically demonstrate that \textit{items share the similar structure of proximity in KG and RS}, thus the cross knowledge transfer of items benefits both recommendation and KGE tasks in MKR.

	\subsection{Results}
    	\subsubsection{Comparison with baselines}
    	\label{sec:observation}
        	The results of all methods in CTR prediction and top-$K$ recommendation are presented in Table \ref{table:ctr} and Figure \ref{fig:precision}, \ref{fig:recall}, respectively.
        	We have the following observations:
        	\begin{itemize}
        		\item
        			PER performs poor on movie, book, and music recommendation because the user-defined meta-paths can hardly be optimal in reality.
        			Moreover, PER cannot be applied to news recommendation.
        		\item
        			CKE performs better in movie, book, and music recommendation than news.
        			This may be because MovieLens-1M, Book-Crossing, and Last.FM are much denser than Bing-News, which is more favorable for the collaborative filtering part in CKE.
        		\item
        			DKN performs best in news recommendation compared with other baselines, but performs worst in other scenarios.
        			This is because movie, book, and musician names are too short and ambiguous to provide useful information.
        		\item
        			RippleNet performs best among all baselines, and even outperforms MKR on MovieLens-1M.
        			This demonstrates that RippleNet can precisely capture user interests, especially in the case where user-item interactions are dense.
        			However, RippleNet is more sensitive to the density of datasets, as it performs worse than MKR in Book-Crossing, Last.FM, and Bing-News.
        			We will further study their performance in sparse scenarios in Section \ref{sec:sparse}.
        		\item
        			In general, our MKR performs best among all methods on the four datasets.
        			Specifically, MKR achieves average $Accuracy$ gains of $11.6\%$, $11.5\%$, $12.7\%$, and $8.7\%$ in movie, book, music, and news recommendation, respectively, which demonstrates the efficacy of the multi-task learning framework in MKR.
        			Note that the top-$K$ metrics are much lower for Bing-News because the number of news is significantly larger than movies, books, and musicians.
        	\end{itemize}
        	
        	\begin{table*}[t]
				\setlength{\tabcolsep}{8pt}
                \centering
                \caption{Results of $AUC$ on MovieLens-1M in CTR prediction with different ratios of training set $r$.}
                \begin{tabular}{c|cccccccccc}
                    \hline
                    \multirow{2}{*}{Model} & \multicolumn{10}{c}{$r$} \\
                    \cline{2-11}
                    & $10\%$ & $20\%$ & $30\%$ & $40\%$ & $50\%$ & $60\%$ & $70\%$ & $80\%$ & $90\%$ & $100\%$ \\
                    \hline
                    PER & 0.598 & 0.607 & 0.621 & 0.638 & 0.647 & 0.662 & 0.675 & 0.688 & 0.697 & 0.710 \\
                    CKE & 0.674 & 0.692 & 0.705 & 0.716 & 0.739 & 0.754 & 0.768 & 0.775 & 0.797 & 0.801 \\
                    DKN & 0.579 & 0.582 & 0.589 & 0.601 & 0.612 & 0.620 & 0.631 & 0.638 & 0.646 & 0.655 \\
                    RippleNet & 0.843 & 0.851 & 0.859 & 0.862 & 0.870 & 0.878 & 0.890 & 0.901 & 0.912 & \textbf{0.920} \\
                    LibFM & 0.801 & 0.810 & 0.816 & 0.829 & 0.837 & 0.850 & 0.864 & 0.875 & 0.886 & 0.892 \\
                    Wide$\&$Deep & 0.788 & 0.802 & 0.809 & 0.815 & 0.821 & 0.840 & 0.858 & 0.876 & 0.884 & 0.898 \\
                    \hline
                    MKR & \textbf{0.868} & \textbf{0.874} & \textbf{0.881} & \textbf{0.882} & \textbf{0.889} & \textbf{0.897} & \textbf{0.903} & \textbf{0.908} & \textbf{0.913} & 0.917 \\
                    \hline
				\end{tabular}
				\label{table:sparse}
			\end{table*}

        \subsubsection{Comparison with MKR variants}
       		We further compare MKR with its three variants to demonstrate the efficacy of cross$\&$compress unit:
       		\begin{itemize}
       			\item MKR-1L is MKR with one layer of cross$\&$compress unit, which corresponds to FM model according to Proposition \ref{prop:1}.
       			Note that MKR-1L is actually MKR in the experiments for MovieLens-1M.
        		\item MKR-DCN is a variant of MKR based on Eq. (\ref{eq:dcn}), which corresponds to DCN model.
        		\item MKR-stitch is another variant of MKR corresponding to the cross-stitch network, in which the transfer weights in Eq. (\ref{eq:cross-stitch}) are replaced by four trainable scalars.
       		\end{itemize}
        
			From Table \ref{table:ctr} we observe that MKR outperforms MKR-1L and MKR-DCN, which shows that modeling high-order interactions between item and entity features is helpful for maintaining decent performance.
			MKR also achieves better scores than MKR-stitch.
			This validates the efficacy of fine-grained control on knowledge transfer in MKR compared with the simple cross-stitch units.
			
		\begin{figure*}[t]
			\centering
            \begin{subfigure}[b]{0.3\textwidth}
                \includegraphics[width=\textwidth]{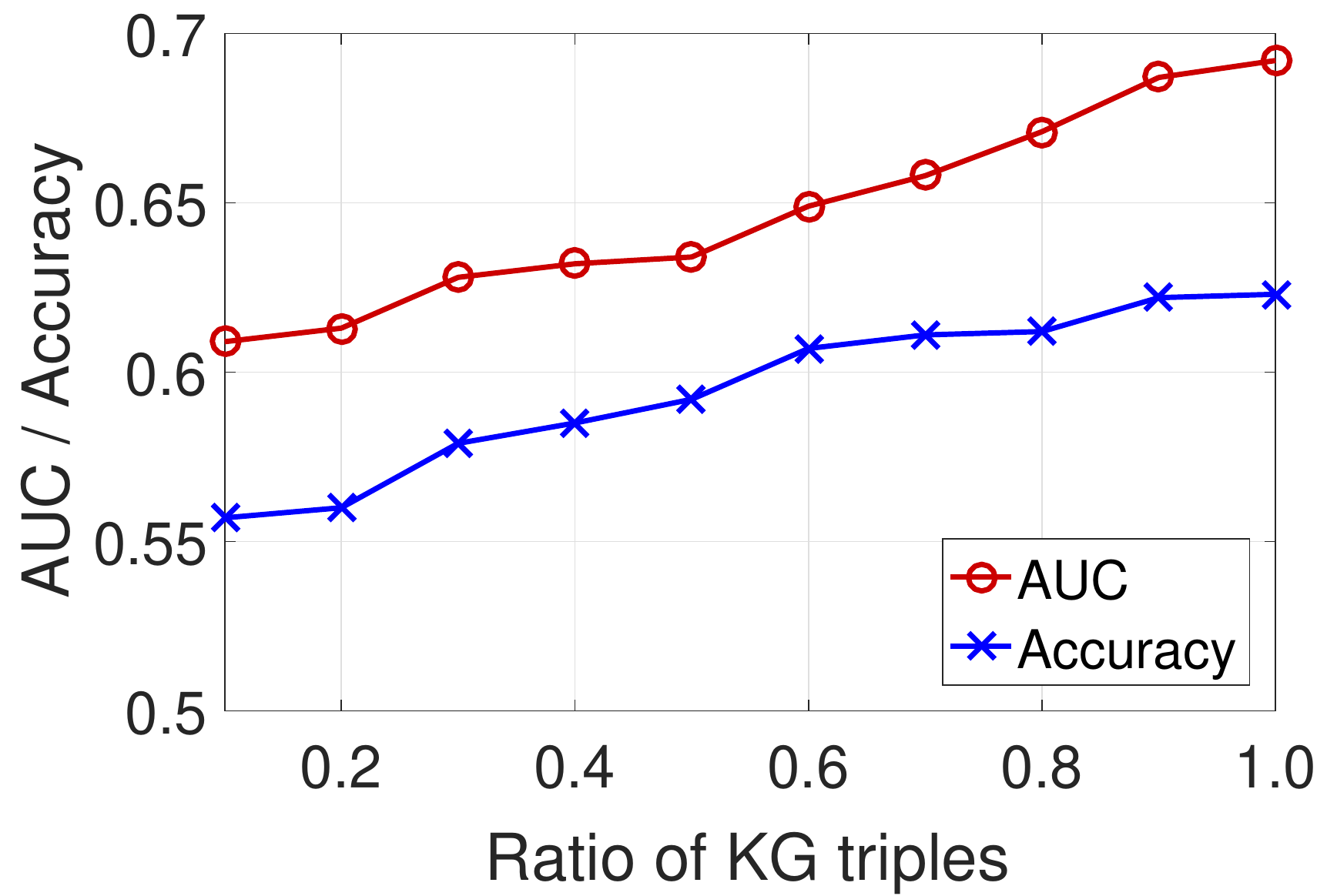}
                \caption{KG size}
                \label{fig:ratio}
            \end{subfigure}
            \hfill
            \begin{subfigure}[b]{0.3\textwidth}
                \includegraphics[width=\textwidth]{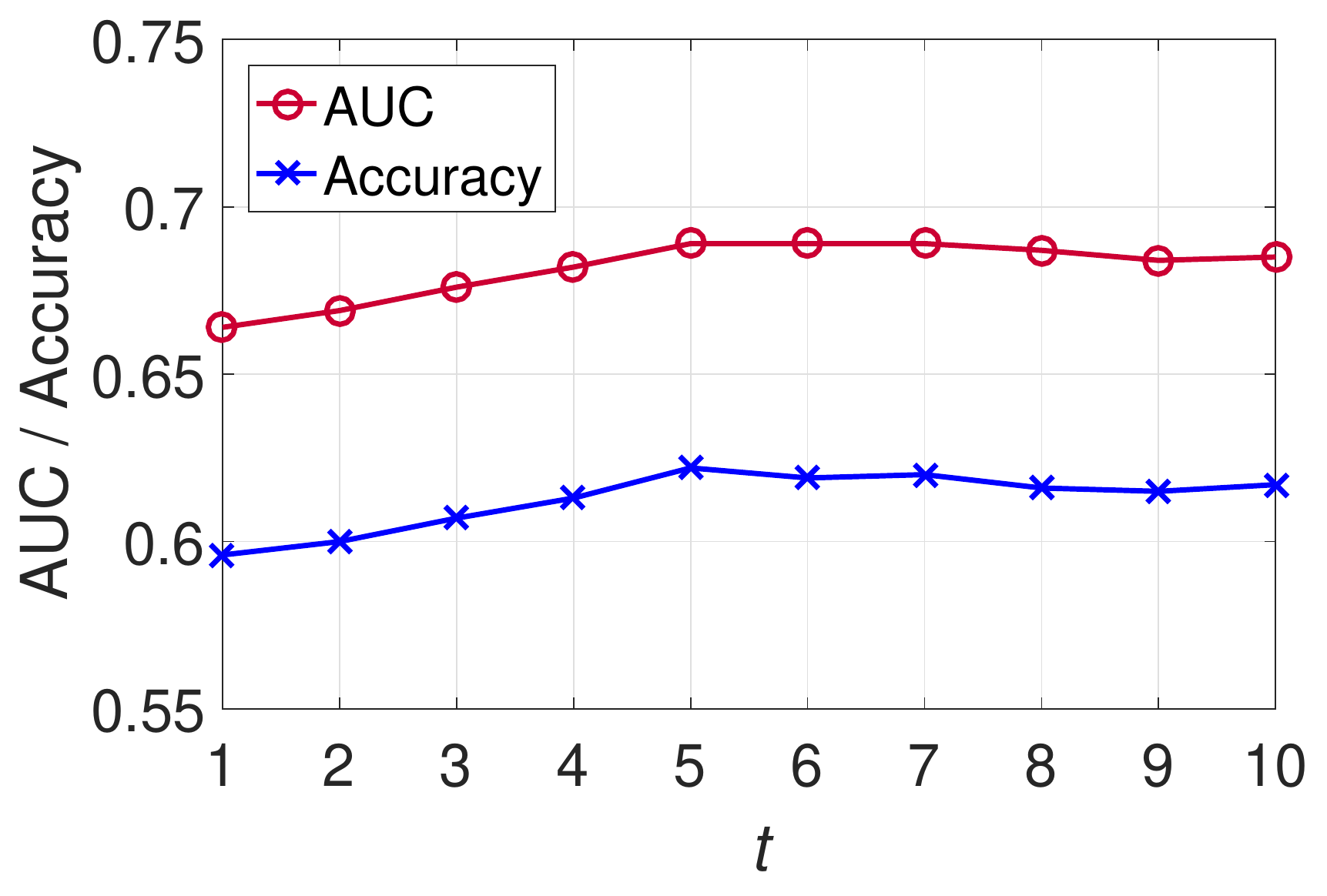}
                \caption{Training frequency}
                \label{fig:t}
            \end{subfigure}
            \hfill
            \begin{subfigure}[b]{0.3\textwidth}
                \includegraphics[width=\textwidth]{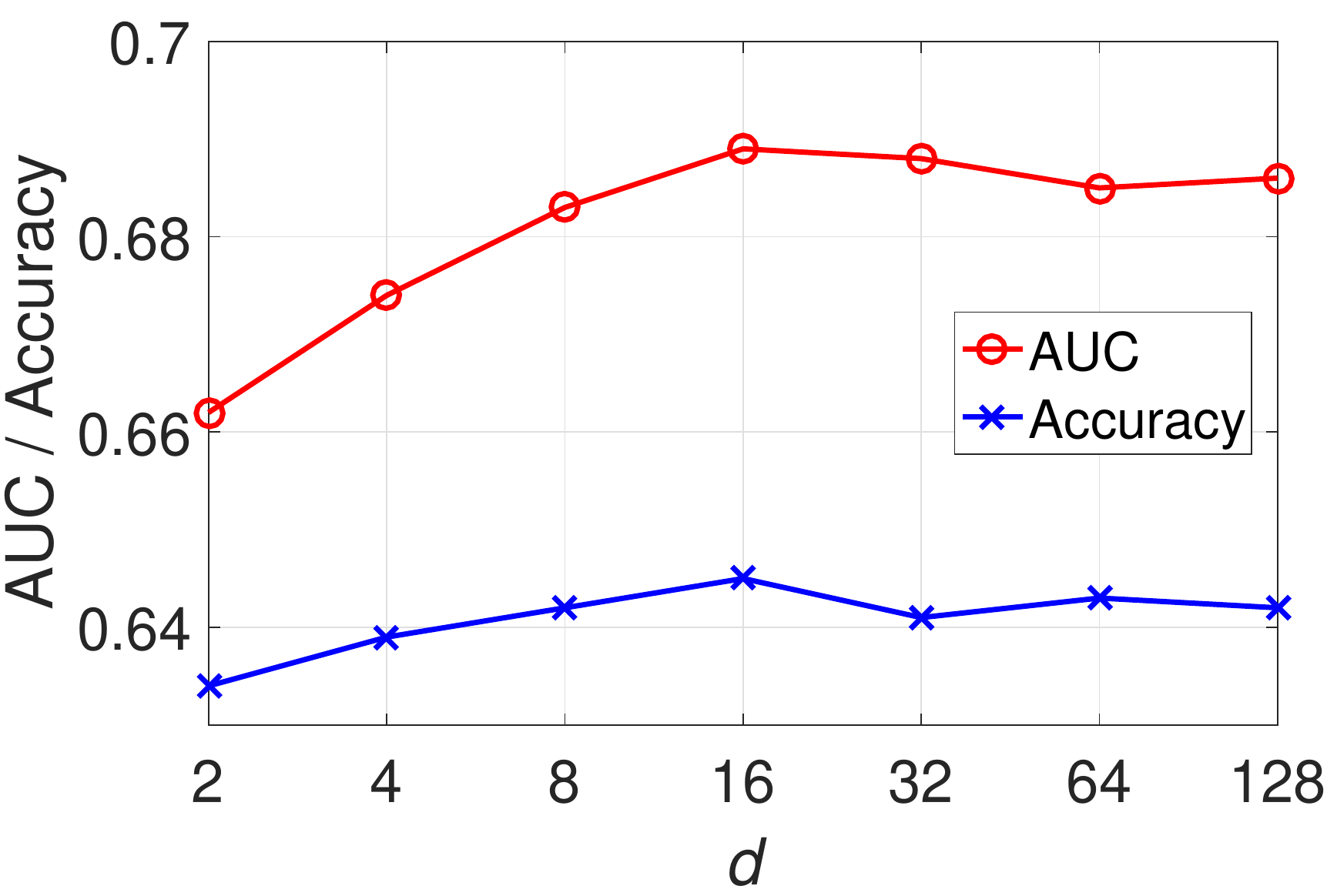}
                \caption{Dimension of embeddings}
                \label{fig:dim}
            \end{subfigure}
            \caption{Parameter sensitivity of MKR on Bing-News w.r.t. (a) the size of the knowledge graph; (b) training frequency of the RS module $t$; and (c) dimension of embeddings $d$.}
        \end{figure*}

       	\begin{table}[t]
			\setlength{\tabcolsep}{10pt}
            \centering
            \caption{The results of $RMSE$ on the KGE module for the three datasets. "KGE" means only KGE module is trained, while "KGE + RS" means KGE module and RS module are trained together.}
            \begin{tabular}{c|cc}
                \hline
                dataset & KGE & KGE + RS \\
                \hline
                MovieLens-1M & 0.319 & \textbf{0.302} \\
                Book-Crossing & 0.596 & \textbf{0.558} \\
                Last.FM & 0.480 & \textbf{0.471}\\
                Bing-News & 0.488 & \textbf{0.459} \\
                \hline
			\end{tabular}
			\label{table:kge}
		\end{table}

		\subsubsection{Results in sparse scenarios}
		\label{sec:sparse}
			One major goal of using knowledge graph in MKR is to alleviate the sparsity and the cold start problem of recommender systems.
			To investigate the efficacy of the KGE module in sparse scenarios, we vary the ratio of training set of MovieLens-1M from $100\%$ to $10\%$ (while the validation and test set are kept fixed), and report the results of $AUC$ in CTR prediction for all methods.
			The results are shown in Table \ref{table:sparse}.
			We observe that the performance of all methods deteriorates with the reduce of the training set.
			When $r=10\%$, the $AUC$ score decreases by $15.8\%$, $15.9\%$, $11.6\%$, $8.4\%$, $10.2\%$, $12.2\%$ for PER, CKE, DKN, RippleNet, LibFM, and Wide$\&$Deep, respectively, compared with the case when full training set is used ($r=100\%$).
			In contrast, the $AUC$ score of MKR only decreases by $5.3\%$, which demonstrates that MKR can still maintain a decent performance even when the user-item interaction is sparse.
			We also notice that MKR performs better than RippleNet in sparse scenarios, which is accordance with our observation in Section \ref{sec:observation} that RippleNet is more sensitive to the density of user-item interactions.

        \subsubsection{Results on KGE side}
        	Although the goal of MKR is to utilize KG to assist with recommendation, it is still interesting to investigate whether the RS task benefits the KGE task, since the principle of multi-task learning is to leverage shared information to help improve the performance of all tasks \cite{zhang2017survey}.
        	We present the result of $RMSE$ (rooted mean square error) between predicted and real vectors of tails in the KGE task in Table \ref{table:kge}.
        	Fortunately, we find that the existence of RS module can indeed reduce the prediction error by $1.9\% \sim 6.4\%$.
       		The results show that the cross$\&$compress units are able to learn general and shared features that mutually benefit both sides of MKR.

	\subsection{Parameter Sensitivity}             
    	\subsubsection{Impact of KG size}
        	We vary the size of KG to further investigate the efficacy of usage of KG.
        	The results of $AUC$ on Bing-News are plotted in Figure \ref{fig:ratio}.
        	Specifically, the $AUC$ and $Accuracy$ is enhanced by $13.6\%$ and $11.8\%$ with the KG ratio increasing from $0.1$ to $1.0$ in three scenarios, respectively.
        	This is because the Bing-News dataset is extremely sparse, making the effect of KG usage rather obvious.

		\subsubsection{Impact of RS training frequency}      
        	We investigate the influence of parameters $t$ in MKR by varying $t$ from 1 to 10, while keeping other parameters fixed.
        	The results are presented in Figure \ref{fig:t}.
        	We observe that MKR achieves the best performance when $t = 5$.
        	This is because a high training frequency of the KGE module will mislead the objective function of MKR, while too small of a training frequency of KGE cannot make full use of the transferred knowledge from the KG.

        \subsubsection{Impact of embedding dimension}
        	We also show how the dimension of users, items, and entities affects the performance of MKR in Figure \ref{fig:dim}.
        	We find that the performance is initially improved with the increase of dimension, because more bits in embedding layer can encode more useful information.
        	However, the performance drops when the dimension further increases, as too large number of dimensions may introduce noises which mislead the subsequent prediction.

\section{Related Work}		
	\subsection{Knowledge Graph Embedding}
		The KGE module in MKR connects to a large body of work in KGE methods.
		KGE is used to embed entities and relations in a knowledge into low-dimensional vector spaces while still preserving the structural information \cite{wang2017knowledge}.
		KGE methods can be classified into the following two categories:
		(1) Translational distance models exploit distance-based scoring functions when learning representations of entities and relations, such as TransE \cite{bordes2013translating}, TransH \cite{wang2014knowledge}, and TransR \cite{lin2015learning};
		(2) Semantic matching models measure plausibility of knowledge triples by matching latent semantics of entities and relations, such as RESCAL \cite{nickel2011three}, ANALOGY \cite{nickel2016holographic}, and HolE \cite{liu2017analogical}.
		Recently, researchers also propose incorporating auxiliary information, such as entity types \cite{xie2016representation}, logic rules \cite{rocktaschel2015injecting}, and textual descriptions \cite{zhong2015aligning} to assist KGE.
		The above KGE methods can also be incorporated into MKR as the implementation of the KGE module, but note that the cross$\&$compress unit in MKR needs to be redesigned accordingly.
		Exploring other designs of KGE module as well as the corresponding bridging unit is also an important direction of future work.

	\subsection{Multi-Task Learning}
		Multi-task learning is a learning paradigm in machine learning and its aim is to leverage useful information contained in multiple related tasks to help improve the generalization performance of all the tasks \cite{zhang2017survey}.
		All of the learning tasks are assumed to be related to each other, and it is found that learning these tasks jointly can lead to performance improvement compared with learning them individually.
		In general, MTL algorithms can be classified into several categories, including feature learning approach \cite{zhang2015deep, wang2017deep}, low-rank approach \cite{han2016multi, mcdonald2014spectral}, task clustering approach \cite{zhou2016flexible}, task relation learning approach \cite{lee2016asymmetric}, and decomposition approach \cite{han2015learning}.
		For example, the cross-stitch network \cite{zhang2015deep} determines the inputs of hidden layers in different tasks by a knowledge transfer matrix;
		Zhou et. al \cite{zhou2016flexible} aims to cluster tasks by identifying representative tasks which are a subset of the given $m$ tasks, i.e., if task $T_i$ is selected by task $T_j$ as a representative task, then it is expected that model parameters for $T_j$ are similar to those of $T_i$.
		MTL can also be combined with other learning paradigms to improve the performance of learning tasks further, including semi-supervised learning, active learning, unsupervised learning,and reinforcement learning.
		
		Our work can be seen as an asymmetric multi-task learning framework \cite{xue2007multi, zhang2012convex, zhang2014regularization}, in which we aim to utilize the connection between RS and KG to help improve their performance, and  the two tasks are trained with different frequencies.

	\subsection{Deep Recommender Systems}
		Recently, deep learning has been revolutionizing recommender systems and achieves better performance in many recommendation scenarios.
		Roughly speaking, deep recommender systems can be classified into two categories:
		(1) Using deep neural networks to process the raw features of users or items \cite{wang2015collaborative, wang2018shine, zhang2016collaborative, wang2017joint, guo2017deepfm};
		For example, Collaborative Deep Learning \cite{wang2015collaborative} designs autoencoders to extract short and dense features from textual input and feeds the features into a collaborative filtering module;
		DeepFM \cite{guo2017deepfm} combines factorization machines for recommendation and deep learning for feature learning in a neural network architecture.
		(2) Using deep neural networks to model the interaction among users and items \cite{huang2013learning, cheng2016wide, covington2016deep, he2017neural}.
		For example, Neural Collaborative Filtering \cite{he2017neural} replaces the inner product with a neural architecture to model the user-item interaction.
		The major difference between these methods and ours is that MKR deploys a multi-task learning framework that utilizes the knowledge from a KG to assist recommendation.

\section{Conclusions and Future Work}
	This paper proposes MKR, a multi-task learning approach for knowledge graph enhanced recommendation.
	MKR is a deep and end-to-end framework that consists of two parts: the recommendation module and the KGE module.
	Both modules adopt multiple nonlinear layers to extract latent features from inputs and fit the complicated interactions of user-item and head-relation pairs. 
	Since the two tasks are not independent but connected by items and entities, we design a cross$\&$compress unit in MKR to associate the two tasks, which can automatically learn high-order interactions of item and entity features and transfer knowledge between the two tasks.
	We conduct extensive experiments in four recommendation scenarios.
	The results demonstrate the significant superiority of MKR over strong baselines and the efficacy of the usage of KG.
	
	For future work, we plan to investigate other types of neural networks (such as CNN) in MKR framework.
	We will also incorporate other KGE methods as the implementation of KGE module in MKR by redesigning the cross$\&$compress unit.

%\section*{Acknowledgments}
%	TODO

\setcounter{theorem}{0}
\setcounter{proposition}{0}
\setcounter{secnumdepth}{4}
\renewcommand\thesubsubsection{\thesubsection.\alph{subsubsection}}

\section*{Appendix}
	\subsection*{A \ \ \ Proof of Theorem \ref{thm:1}}
	
	\begin{proof}
		We prove the theorem by induction:
		
		\textbf{Base case}: When $l=1$,
		\begin{equation*}
		\begin{split}
			{\bf v}_1 =& {\bf v} {\bf e}^\top {\bf w}_0^{VV} + {\bf e} {\bf v}^\top {\bf w}_0^{EV} + {\bf b}_0^V\\
			=& \left[v_1 \sum_{i=1}^d e_i w_0^{VV(i)} \ \cdots \ v_d \sum_{i=1}^d e_i w_0^{VV(i)} \right]^\top\\
			&+ \left[e_1 \sum_{i=1}^d v_i w_0^{EV(i)} \ \cdots \ e_d \sum_{i=1}^d v_i w_0^{EV(i)} \right]^\top\\
			&+ \left[ b_0^{V(0)} \ \cdots \ b_0^{V(d)} \right]^\top.
		\end{split}
		\end{equation*}
		
		Therefore, we have		
		\begin{equation*}
		\begin{split}
			\|{\bf v}_1\|_1 =& \left| \sum_{j=1}^d v_j \sum_{i=1}^d e_i w_0^{VV(i)} + \sum_{j=1}^d e_j \sum_{i=1}^d v_i w_0^{EV(i)} + \sum_{i=1}^d b_0^{V(d)} \right|\\
			=& \left| \sum_{i=1}^d \sum_{j=1}^d (w_0^{EV(i)} + w_0^{VV(j)}) v_i e_j + \sum_{i=1}^d b_0^{V(d)} \right|.
		\end{split}
		\end{equation*}
		
		It is clear that the cross terms about ${\bf v}$ and ${\bf e}$ with maximal degree is $k_{\bm \alpha, \bm \beta} v_i e_j$, so we have $\alpha_1 + \cdots + \alpha_d = 1 = 2^{1-1}$, and $\beta_1 + \cdots + \beta_d = 1 = 2^{1-1}$ for ${\bf v}_1$.
		The proof for ${\bf e}_1$ is similar.
		
		\textbf{Induction step}: Suppose $\alpha_1 + \cdots + \alpha_d = 2^{l-1}$ and $\beta_1 + \cdots + \beta_d = 2^{l-1}$ hold for the maximal-degree term $x$ and $y$ in $\|{\bf v}_l\|_1$ and $\|{\bf e}_l\|_1$.
		Since $\|{\bf v}_l\|_1 = \left| \sum_{i=1}^d v_l^{(i)} \right|$ and $\|{\bf e}_l\|_1 = \left| \sum_{i=1}^d e_l^{(i)} \right|$, without loss of generosity, we assume that $x$ and $y$ exist in $v_l^{(a)}$ and $e_l^{(b)}$, respectively.
		Then for $l+1$, we have
		\begin{equation*}
			\|{\bf v}_{l+1}\|_1 = \sum_{i=1}^d \sum_{j=1}^d (w_l^{EV(i)} + w_l^{VV(j)}) v_l^{(i)} e_l^{(j)} + \sum_{i=1}^d b_l^{V(d)}.
		\end{equation*}
		
		Obviously, the maximal-degree term in $\|{\bf v}_{l+1}\|_1$ is the cross term $xy$ in $v_l^{(a)} e_l^{(b)}$.
		Since we have $\alpha_1 + \cdots + \alpha_d = 2^{l-1}$ and $\beta_1 + \cdots + \beta_d = 2^{l-1}$ for both $x$ and $y$, the degree of cross term $xy$ therefore satisfies $\alpha_1 + \cdots + \alpha_d = 2^{(l+1)-1}$ and $\beta_1 + \cdots + \beta_d = 2^{(l+1)-1}$.
		The proof for $\|{\bf e}_{l+1}\|_1$ is similar.
	\end{proof}

	\subsection*{B \ \ \ Proof of Proposition \ref{prop:1}}
	
	\begin{proof}
		In the proof of Theorem \ref{thm:1} in Appendix A, we have shown that
		\begin{equation*}
			\|{\bf v}_1\|_1 = \left| \sum_{i=1}^d \sum_{j=1}^d (w_0^{EV(i)} + w_0^{VV(j)}) v_i e_j + \sum_{i=1}^d b_0^{V(d)} \right|.
		\end{equation*}
		
		It is easy to see that $w_i = w_0^{EV(i)}$, $w_j = w_0^{VV(j)}$, and $b = \sum_{i=1}^d b_0^{V(d)}$.
		The proof is similar for $\|{\bf e}_1\|_1$.
	\end{proof}
	
	\vspace{1em}
	
	We omit the proofs for Proposition \ref{prop:2} and Proposition \ref{prop:3} as they are straightforward.

\bibliographystyle{ACM-Reference-Format}
\bibliography{reference}

\end{document}